\documentclass[fleqn,usenatbib]{mnras}

\usepackage{newtxtext,newtxmath}

\usepackage[T1]{fontenc}
\usepackage{textcomp}
\usepackage{xspace}
\usepackage{mathpazo} 
\usepackage{amsmath}
\usepackage{esint}
\usepackage{adjustbox}
\usepackage{rotating}
\usepackage{tablefootnote} 
\usepackage{siunitx}
\usepackage{subfig}
\DeclareRobustCommand{\VAN}[3]{#2}
\let\VANthebibliography\thebibliography
\def\thebibliography{\DeclareRobustCommand{\VAN}[3]{##3}\VANthebibliography}

\usepackage{graphicx}	
\usepackage{amsmath}	
\usepackage{amssymb}	
\usepackage[T1]{fontenc}
\usepackage{soul}

\newcommand{\hi}{{\sc H\,i}\xspace} 

\newcommand{\kms}{$\,$km$\,$s$^{-1}$}

\newcommand{\perbeam}{beam$^{-1}$}

\title[HI in the Frontier Fields]{HI in and behind the Hubble Frontier Field Clusters: A Deep MeerKAT Pilot Search out to $z\sim0.5$}

\author[S. Ranchod et al.]{
Shilpa Ranchod$^{1,2}$\thanks{E-mail: shilparanchod@gmail.com},
Roger Deane$^{1,2}$,
Danail Obreschkow$^{3}$,
Tariq Blecher$^{4}$,
Ian Heywood$^{4,5,6}$,
\\
$^{1}$Department of Physics, University of Pretoria, Private Bag X20, Pretoria 0028, South Africa\\
$^{2}$Wits Centre for Astrophysics, School of Physics, University of the Witwatersrand, 1 Jan Smuts Avenue, 2000, South Africa \\
$^{3}$International Centre for Radio Astronomy Research (ICRAR), M468, University of Western Australia, WA 6009, Australia \\
$^{4}$Department of Physics and Electronics, Rhodes University, PO Box 94, Makhanda 6140, Eastern Cape, South Africa\\
$^{5}$Oxford Astrophysics, Denys Wilkinson Building, University of Oxford, Keble Rd, Oxford, OX1 3RH, UK \\
$^{6}$South African Radio Astronomy Observatory, 2 Fir Street, Observatory, 7925, South Africa
}

\date{Accepted XXX. Received YYY; in original form ZZZ}

\pubyear{2015}

\begin{document}
\label{firstpage}
\pagerange{\pageref{firstpage}--\pageref{lastpage}}
\maketitle

\begin{abstract}
The Hubble Frontier Fields (HFF) are a selection of well-studied galaxy clusters used to probe dense environments and distant gravitationally lensed galaxies. We explore the 21cm neutral hydrogen (\hi) content of galaxies in three of the HFF clusters, Abell~2744 ($z$ = 0.308), Abell~S1063 ($z$ = 0.346) and Abell~370 ($z$ = 0.375), to investigate the evolution of gas in galaxies within intermediate redshift clusters. 
Using Early Science MeerKAT observations, we perform spectral-line stacking with \hi cubes and make a 3$\sigma$ stacked detection for blue galaxies in Abell~S1063 ($M_\mathrm{HI} = 1.22^{+0.38}_{-0.36}\,\times 10^{10}\,\mathrm{M}_\odot$). We determine the 3$\sigma$ \hi mass detection limits of Abell~2744 and Abell~370 to be at the knee of the \hi Mass Function. A final, more ambitious objective of this work is to search for gravitationally lensed \hi emission behind these clusters, enabled by MeerKAT's wide instantaneous bandwidth. We find no evidence of highly magnified \hi emission at $0.33<z<0.58$. The low thermal noise levels achieved in these pilot observations, despite short integration times, highlights the enormous potential of future MeerKAT \hi observations of dense environments and the intermediate-redshift Universe.
\end{abstract}





\section{Introduction}\label{ssec:4-intro}

Environment has been shown to play an important role in galaxy evolution. In dense environments, such as galaxy clusters and groups, galaxies are subject to a multitude of effects that can alter their morphology, gas and star-formation properties \citep[e.g.][]{Verheijen_2004,D_nes_2015}. A statistical result of this, the morphology-density relation \citep[e.g.][]{Dressler_1997,van_der_Wel_2010,Wetzel_2013}, shows that early-type (elliptical and lenticular) galaxies are preferably found in clusters, with the number of late-type (spiral and irregular) galaxies decreasing with environment density in the local Universe. 

Galaxy clusters are the most massive virialised structures in the Universe and grow via the accretion of gas and galaxies along filaments in the cosmic web \citep[e.g.][]{Salerno_2020}. 
Within galaxy clusters, an important transformation mechanism is ram pressure stripping \citep{Gunn_1972}, where the gas of the interstellar medium (ISM) of an infalling galaxy is stripped by the intracluster medium \citep[ICM, e.g.][]{Vollmer_2013,Sorgho_2017,Ramatsoku_2020}. In addition to interactions with the ICM, cluster galaxies are subject to fast interactions with other cluster members, known as galaxy harassment \citep{Moore_1998,Bialas_2015}. These processes can result in gas being removed from galaxies, or rapidly increases the star-formation rate, which quickly depletes the gas reservoir \citep[$\sim\,10^7$ years,][]{Poggianti_2004}. Because gas serves as the essential fuel for star-formation, these mechanisms are directly linked to the quenching of star-formation in the cluster environment, which results in the reddening of galaxies and the transformation of morphology \citep[e.g.][]{Odekon_2016,Joshi_2020}.  

A fundamental tool for studying these gas removal processes is the observation of neutral hydrogen (\hi). \hi makes up a large component of the ISM and can be directly observed via the 21~cm emission line. It has an important role in the baryon cycle, in which it is converted into H$_2$, which fuels star-formation \citep{Keres_2005,Obreschkow_2009,P_roux_2020}. Within galaxies, \hi is diffuse, and extends beyond the stellar component \citep[e.g.][]{Broeils_1997,Leroy_2008}, making it a sensitive long dynamical tracer of the different environmental processes \citep[e.g.][]{Oosterloo_2005,Serra_2012b,Saponara_2017}. 
It is crucial to study the \hi content and distribution in cluster galaxies, to fully understand these mechanisms \citep[e.g.][]{Brown_2016,Deshev_2020}. Such observations have shown that the effect of ram-pressure stripping is the strongest in the central regions of dense clusters, with evidence showing that the \hi can be fully depleted after a galaxy has passed through the cluster centre \citep{Roediger_2007}. Observations of low-redshift clusters have shown unusual disruptions in the \hi-discs of cluster galaxies, tidal features, and \hi deficient galaxies \citep[e.g.][]{Solanes_2001,Jaffe_2015,Lee_Waddell_2017,Gavazzi_2018}. Some galaxies are thought to go through environment-induced ``pre-processing'' before entering a massive cluster environment \citep{Li_2020}. The pre-processing takes place in low-mass groups in the form of galaxy-galaxy interactions and ram pressure stripping \citep{Hess_2013,Jaffe_2013,Jaffe_2015}, which typically reduces the \hi content of these galaxies. Tracing the pre-processing is important for understanding the gas content of galaxies, and how this evolves with large-scale structure and over cosmic time. 

\hi has not been observed in detail beyond $z=0.2$, due to the faintness of the \hi emission line, with the record for the highest redshift direct \hi emission detection of a single galaxy at $z = 0.37$, with $M_\mathrm{HI} = 2.9 \times 10^{10} \mathrm{M}_\odot$ \citep{Fern_ndez_2016}. This galaxy resides beyond the knee of the HIMF and is therefore not representative of the general \hi galaxy population at this redshift, however, statistical detections of intermediate redshift \hi are possible. \hi spectral stacking has been highly successful in quantifying the average \hi content of large numbers of galaxies in the nearby Universe, and beyond $z = 0.2$ \citep[e.g.][]{Delhaize_2013,Rhee_2013,Kanekar_2016,Bera_2019}, with the most distant stacked \hi detection at $0.74 < z < 1.45$ for 7653 blue, star-forming galaxies \citep{Chowdhury_2020}. With this technique, the average \hi mass of a region (e.g. a galaxy cluster, group or blind field) is determined by co-adding the appropriately shifted spectra for a sample of galaxies in the region. 
\hi spectral stacking has been successful in detecting the average \hi in and around $0.2 < z < 0.3$ galaxy clusters \citep{Verheijen_2007,Lah_2009}, and in the local Universe, has been used to probe the \hi substructure of the Coma cluster \citep{Healy_2021}.

In addition to being useful laboratories for studying the evolution of galaxies in dense environments, galaxy clusters can be 
effective gravitational lenses \citep[for a review, see][]{Kneib_2011}. Gravitational lensing is the deflection of light by intervening mass, that can produce highly magnified and distorted images of background galaxies. The large mass and solid angle covered by highly concentrated galaxy clusters make them ideal gravitational lenses, 
which can be used as ``cosmic telescopes'' to observe very distant galaxies \citep[e.g.][]{Richard_2009}. 
The amplification of sources through gravitational lensing has been extremely effective in observing faint, distant sources across the electromagnetic spectrum, including continuum and spectral line emission in the radio domain \citep[e.g.][and references therein]{Carilli_2013}. While molecular gas has been studied across the Universe up to $z \sim 1$ and beyond, \hi emission remains undetected through lensing, with two searches for galaxy-galaxy lensed \hi sources at $z \sim 0.4$ \citep{Hunt_2016,Blecher_2019}. 
Gravitational lensing conserves surface brightness while increasing the solid angle of the source, boosting the observed flux. This amplification $\mu$ can facilitate the detection of unresolved lensed sources, which maximises their detection probability, and reduces the integration time needed for a given source by $\mu^2$. 
Next-generation cm-wavelength interferometers are now sensitive enough to 
observe the higher redshift \hi 
Universe, and the detection of gravitationally lensed \hi is probable in new surveys \citep{Deane_2015}. The detection of lensed \hi behind intermediate-redshift galaxy clusters will provide a deep 
cosmic view of \hi emission in galaxies, pre-SKA era, within a fraction of the observation time of unlensed detections. Successful lensed \hi detections, along with readily detected CO emission lines, will constrain the \hi/$\mathrm{H}_2$ ratio at these redshifts, an important parameter in understanding galaxy evolution over cosmic time \citep{Obreschkow_2009}. 

MeerKAT's wide instantaneous bandwidth enables us to search for \hi within intermediate-redshift galaxy clusters, and gravitationally lensed \hi behind certain clusters in the L-band. 
Its large field of view makes it possible to observe these clusters with a single pointing and with shorter integration times than previous generation telescopes. Surveys such as LADUMA and MIGHTEE plan to detect \hi emission up to a redshift of $z\sim0.58$ in the L-band and $z\sim1.4$ in the UHF-band \citep{Jarvis_2017,Blyth_2016}. Direct \hi emission detections from these surveys are limited to high mass ($M_\mathrm{HI} > 10^{10}\,\mathrm{M}_\odot$) galaxies beyond $z\sim0.35$, with lower \hi mass detections possible through statistical methods, such as stacking. 
The MeerKAT Galaxy Cluster Legacy Survey \citep[MGCLS;][]{Knowles_2021}, is the MeerKAT L-band survey of 115 galaxy clusters. This survey includes the observations of three of the Hubble Frontier Field (HFF) Clusters: Abell~2744, Abell~S1063 and Abell~370 ($0.3 < z < 0.4$). These massive clusters have strong lensing capabilities, which have been studied and modelled in detail with optical and infrared (OIR) observations \citep[e.g.][]{Jauzac_2015,Mahler_2017}.

In this paper, we perform a deep search for \hi emission in three of the southern/equatorial HFF clusters with MeerKAT, through direct detections and \hi spectral stacking. In addition, we search for gravitationally lensed \hi behind these clusters, up to $z = 0.58$. The paper is structured as follows. In Section~\ref{sec:4-data} we describe the MeerKAT observations and data calibration strategy and summarise the ancillary data. In Section~\ref{sec:4-results} we present the source-finding and stacking results. In Section~\ref{sec:gravlens} we show results for the lensed \hi search. Section~\ref{sec:4-discussion} is a discussion of the results, and a summary follows in Section~\ref{sec:4-conclusion}.

Throughout this work we assume cosmological values of $\Omega_M=0.307, \Omega_\Lambda=0.691$ and $H_0 = 67.7\,$\kms $\mathrm{Mpc}^{-1}$ \citep{Planckcollab_2016}.

\section{Data}\label{sec:4-data}

\subsection{Hubble Frontier Field Clusters}
Of the six HFF clusters, four are in the Southern/Equatorial sky and are therefore 
more conducive to deep, high fidelity imaging with MeerKAT. These clusters, namely Abell~2744, Abell~S1063, Abell~370 and MACS~J0416 are in the redshift range of $0.3 < z < 0.4$, and are some of the most massive clusters at these redshifts \citep{Lotz_2017}. This range provides a balance between low-redshift, where clusters have low surface mass density, and high-redshift, where cluster mass has not significantly built up, as well as avoiding the redshift range with high RFI ($0.1<z<0.3$). In addition, the cluster redshift range provides the optimal lensing efficiency for sources within the MeerKAT L- and UHF-bands. In this work, we observe Abell~2744, Abell~S1063 and Abell~370, with plans for future observations of MACS~J0416. The properties of the three observed clusters are summarised in Table~\ref{tab:4-cluster} \citep[e.g.][]{Dressler_1999,Owers_2011,Williamson_2011}.

\begin{table}
\centering
\small
\caption{Cluster properties. The far-right column only includes cluster members with spectroscopic redshifts.}
\begin{adjustbox}{width=0.5\textwidth,center}
\begin{tabular}{cccccc}
\hline
            & RA (J2000) & Dec (J2000) & $z$    & $\sigma$                   &  No.       \\
            &     &    &        & [\kms] & members \\ \hline
Abell~2744  & 00$^h$14$^m$21.$^s$2 & -30$^\circ$23'50."1 & 0.308  & $1497 \pm 47$             & 167    \\
Abell~S1063 & 22$^h$48$^m$44.$^s$4 & -44$^\circ$31'48."5 & 0.346 & $1840^{+230}_{-150}$      & 106     \\
Abell~370   & 02$^h$39$^m$52.$^s$9 & -01$^\circ$34'36."5 & 0.375  & $\sim1170$                & 50     \\ \hline
\end{tabular}
\end{adjustbox}
\label{tab:4-cluster}
\end{table}

The HFF clusters have been well studied in a large range of wavelengths, including some of the deepest OIR observations of clusters (see Table~\ref{tab:4-cluster}), but have yet to be observed in \hi emission, except for Abell~370 \citep{Lah_2009}. Each cluster has an extensive suite of multi-wavelength catalogues \citep{Shipley_2018}, including a fairly complete sample with spectroscopic redshifts \citep{Owers_2011,Karman_2015,Mahler_2017,Lagattuta_2017}. The precision of these redshift measurements is necessary for targeted \hi surveys and statistical detection methods - i.e. stacking. The HFF clusters also have accurate mass models available from OIR lensing and X-ray analyses \citep[e.g.][]{Jauzac_2015}, an important tool needed to identify and model lensed \hi detections. 

\subsection{MeerKAT Observations}\label{ssec:4-meerkatobs}
The MeerKAT Galaxy Cluster Legacy Survey is a deep survey of 115 legacy galaxy clusters. The majority of observations make use of more than 60 antennas and consist of $\sim10$ hour integration times. The observations used the 4K correlator mode which has a channel width of 209\,kHz, corresponding to 44\kms at $z=0$. The complex gain and bandpass calibrators, as well as additional information for each observation, are summarised in Table~\ref{tab:4-obsandcal}.
\begin{table}
\footnotesize
\centering
\caption[Observation and Calibration Details for MeerKAT data]{Observation and Calibration Details for MeerKAT data}
\begin{adjustbox}{width=0.5\textwidth,center}
\begin{tabular}{lccc}
\hline
                                  & Abell~2744      & Abell~S1063     & Abell~370     \\ \hline
RA (J2000)         & 00$^h$14$^m$19.$^s$51   & 22$^h$48$^m$43.$^s$50   & 02$^h$39$^m$50$^s$.50   \\
Dec (J2000)        & -30$^\circ$23'19."20  & -44$^\circ$31'44."00  & -01$^\circ$35'08."00  \\
Date                              & 2018 July 5     & 2018 July 3     &  2019 June 16 \\
Time range (UTC) & 00:17 - 06:33 & 21:27 - 06:40 & 04:18 - 11:58 \\
Number of antennas                & 61              & 61              &   62        \\
Bandpass/flux calibrator          & J0408-6545        & J1939-6342    & J0408-6545    \\
Gain calibrator                   & J0025-2602        & J2314-4455      &  J0323+0534   \\
Time on target                    & 6.27 hr         & 6.74 hr          & 7.27 hr    \\
\noalign{\vskip .06in}
\hline
\noalign{\vskip .06in}   
Processed frequency range &  946--1092 MHz &   856--1712 MHz &  898--1423 MHz     \\
Continuum Image size               &  $5000\times5000$ &   $10240\times10240$ &  $10240\times10240$             \\
Briggs weighting         &  0.0   &   -0.3      &   -0.3            \\
Pixel Scale       &  2.0" &  1.1"            &  1.1"             \\
Restoring beam FWHM      &$9.8"\times 8.7"$ &  $6.2"\times5.9"$ &  $8.1"\times 6.1"$ \\
Image rms            & 9$\,\mu$Jy \perbeam  &    3$\,\mu$Jy \perbeam   &    5$\,\mu$Jy \perbeam           \\ \hline
\hline
\end{tabular}
\end{adjustbox}
\label{tab:4-obsandcal}
\end{table}
%
\subsection{Calibration and Imaging}\label{sssec:4-meerkatcal}
%
The MeerKAT data were calibrated (1GC and 2GC) and imaged using the {\sc Oxkat}\footnote{\url{https://github.com/IanHeywood/oxkat}} \citep{Heywood_2020} and {\sc Caracal}\footnote{\url{https://github.com/caracal-pipeline/caracal}} \citep{Jozsa_2020} pipelines. 
{\sc Oxkat} is composed of a suite of packages, including {\sc Casa} {\citep{McMullin_2007}}, {\sc Wsclean} \citep{Offringa_2014} and the \textsc{Tricolour}\footnote{\url{https://github.com/ska-sa/tricolour}} flagging software. The {\sc Oxkat} scripts were run within {\sc Singularity} containers on the ILIFU Cloud facility\footnote{\url{http://www.ilifu.ac.za/}}, and the pipeline follows a standard calibration strategy. Briefly, the calibration process is as follows: 

\begin{enumerate}
    \item {\bf 1GC:} Basic flagging is applied to the calibrator fields, and auto-flagging on model-subtracted calibrators. These flags are then applied to the target. Flux, delay, bandpass and gain calibrations are derived and applied iteratively, with rounds of flagging in between. The target data are split out of the Measurement Set. These tasks are executed with {\sc Casa}, and diagnostic visibility plots are generated with \textsc{Ragavi}\footnote{\url{https://github.com/ratt-ru/ragavi/}} (Radio Astronomy Gains And Visibility Inspector) and \textsc{shadeMS}\footnote{\url{https://github.com/ratt-ru/shadeMS}}, a tool for plotting interferometric visibilities.
    \item {\bf Imaging:} The target data are flagged using {\sc Tricolour}, and imaged using {\sc Wsclean}. The initial imaging run is blind and shallow, with imaging parameters as summarised in Table~\ref{tab:4-obsandcal}. A deconvolution mask is then created from the image using local RMS thresholding.
    \item {\bf 2GC:} Masked deconvolution is performed using {\sc Wsclean}, from which model visibilities are predicted using {\sc Wsclean}. One round of phase-only self-calibration is executed using {\sc Casa} tasks\footnote{A more recent version of {\sc Oxkat} now uses {\sc CubiCal} \citep{Kenyon_2018} and performs phase and delay self-calibration.}, and the result is imaged using {\sc Wsclean}. Imaging parameters are summarised in Table~\ref{tab:4-obsandcal}.
\end{enumerate}
Since this work began in the early stages of MeerKAT science verification, multiple approaches were used for calibration. {\sc Caracal} is a radio calibration pipeline that makes use of the {\sc Stimela}\footnote{\url{https://github.com/ratt-ru/Stimela}} \citep{Makhathini_2018} scripting framework. {\sc Stimela} is a Python- and container-based framework that allows users to execute tasks from different radio calibration software packages using a single configuration file. These packages include {\sc Casa}, {\sc MeqTrees} \footnote{\url{https://github.com/ska-sa/meqtrees}} \citep{Noordam_2010}, and {\sc SoFiA} \citep{Serra_2015}, among others. {\sc Caracal} was also run on the ILIFU Cloud facility. The calibration strategy is as follows:
\begin{enumerate}
    \item {\bf 1GC:} Before cross-calibration, basic flagging on auto-correlations are applied to all fields using the {\sc Casa} task \texttt{flagdata}, and autoflagging of RFI is done using {\sc Aoflagger} \citep{Offringa_2010}. As done with {\sc Oxkat}, flux, delay, bandpass and gain calibrations are derived and applied iteratively, with rounds of flagging in between, using {\sc Casa}, and the target data were split out of the MS.
    \item {\bf Flagging:} The target MS is autoflagged with {\sc Aoflagger}, using a custom flagging strategy.
    \item {\bf Imaging and 2GC}: This pipeline produces five images using {\sc WSClean}, with four rounds of self-calibration using {\sc CubiCal}\footnote{\url{https://github.com/ratt-ru/CubiCal}} \citep{Kenyon_2018}. The {\sc WSClean} \texttt{auto-mask} and \texttt{auto-threshold} parameters range from 20 to 5, and 0.5 to 0.3 respectively, decreasing with iterations. The other imaging parameters are summarised in Table~\ref{tab:4-obsandcal}. The first three rounds of self-calibration are phase-only, and the final round calibrates both phase and amplitude.
\end{enumerate}
The data for Abell~2744 were calibrated using {\sc Caracal}, and Abell~S1063 and Abell~370 were calibrated using {\sc Oxkat}. For Abell~370, we did an additional round of manual flagging of short baselines (< 300m) before 1GC using {\sc Aoflagger}, following inspection.
\begin{table*}
\centering
\caption[Cluster-centred \hi cube imaging parameters and cube properties.]{\hi cube imaging parameters and cube properties.}
\begin{tabular}{lccc}
\hline
                     & Abell~2744 & Abell~S1063 & Abell~370 \\ \hline
\underline{Cluster cube:}                     &  &   &   \\ 
Cube frequency range & $1072 - 1078$ MHz & $1049 - 1068$ MHz & $1020-1045$ MHz         \\
Cube redshift range  & $0.300 - 0.317$ & $0.329 - 0.354$ &  $0.359-0.392$         \\
Cube velocity range  & 72318 - 73585 \kms & 74430 - 78443 \kms & 79288 - 84568 \kms \\
Image size (pixels)          & 1000$\times$1000 & 1000$\times$1000& $2800\times2800$          \\
Image size (Mpc)    & 8.7$\times$ 8.7 & 9.0$\times$ 9.0 & 9.6$\times$ 9.6  \\
Pixel scale   & 2.0"         & 2.0"            &  2.5"         \\
Briggs weighting     & 0.5        & 0.5            & 0.5           \\
Restoring beam FWHM  & 17.0" $\times$ 11.3" & 14.8" $\times$ 12.7" & 23.6" $\times$ 13.2"          \\
Median channel map rms  & 112$\,\mu$Jy \perbeam   & 107$\,\mu$Jy \perbeam   & 124$\,\mu$Jy \perbeam \\ 
\noalign{\vskip .03in}
\hline
\noalign{\vskip .03in} 
\underline{Background cube:} &  &   & \\
Cube frequency range & 974-1026 MHz  &  963-1046 MHz     & 966-1022 MHz \\
Cube redshift range  &  0.385 - 0.459    & 0.357 - 0.474     & 0.390 - 0.470       \\
Image size (pixels)           & $1000\times1000$ & $1000\times1000$  & $2800\times2800$ \\
Max image size (Mpc)    & 10.2$\times$ 10.2 & 10.9$\times$ 10.9 & 10.8$\times$ 10.8  \\
Pixel scale   & 2.0"             & 2.0"              & 2.5"   \\
Briggs weighting     &  0.5             & 0.5               & 0.5    \\
Restoring beam FWHM  & $17.3"\times11.7"$  & $15.4"\times13.5"$  & $22.6"\times12.6"$          \\
Median cube rms      & 113 $\mu$Jy \perbeam        & 110 $\mu$Jy \perbeam         &  129 $\mu$Jy \perbeam   \\ \hline
\end{tabular}
\label{tab:4-cubeparameters}
\end{table*}

After assessing that the quality of the calibration was satisfactory for \hi imaging, the model visibilities were subtracted using {\sc Casa's} \texttt{uvsub} task. For each cluster, spectral line imaging was performed for all the channels with frequencies corresponding to the redshift of the cluster and behind the cluster up to $z<0.5$. The low-frequency end of the L-band (full bandwidth extent corresponds to $z < 0.58$) was omitted, due to high RFI contamination at these frequencies. The Briggs \texttt{robust} weighting parameter was chosen for maximum sensitivity to faint emission. `Cluster' cubes of $\sim 100$ channels were constructed, centred around each cluster frequency, with channel widths corresponding to $\sim 60$\kms at $z=0.35$ and `background' cubes were generated for lower frequencies corresponding up to $z<0.5$. The selected frequency ranges for each cluster and the imaging parameters, along with the resultant cube PSF and rms values are summarised in Table~\ref{tab:4-cubeparameters}. The {\sc Casa} task {\tt imcontsub} was used to remove residual continuum emission by subtracting a first or second-order polynomial from the spectral axis of the \hi cubes. 

\begin{figure}
    \centering
    \includegraphics[width=0.99\columnwidth]{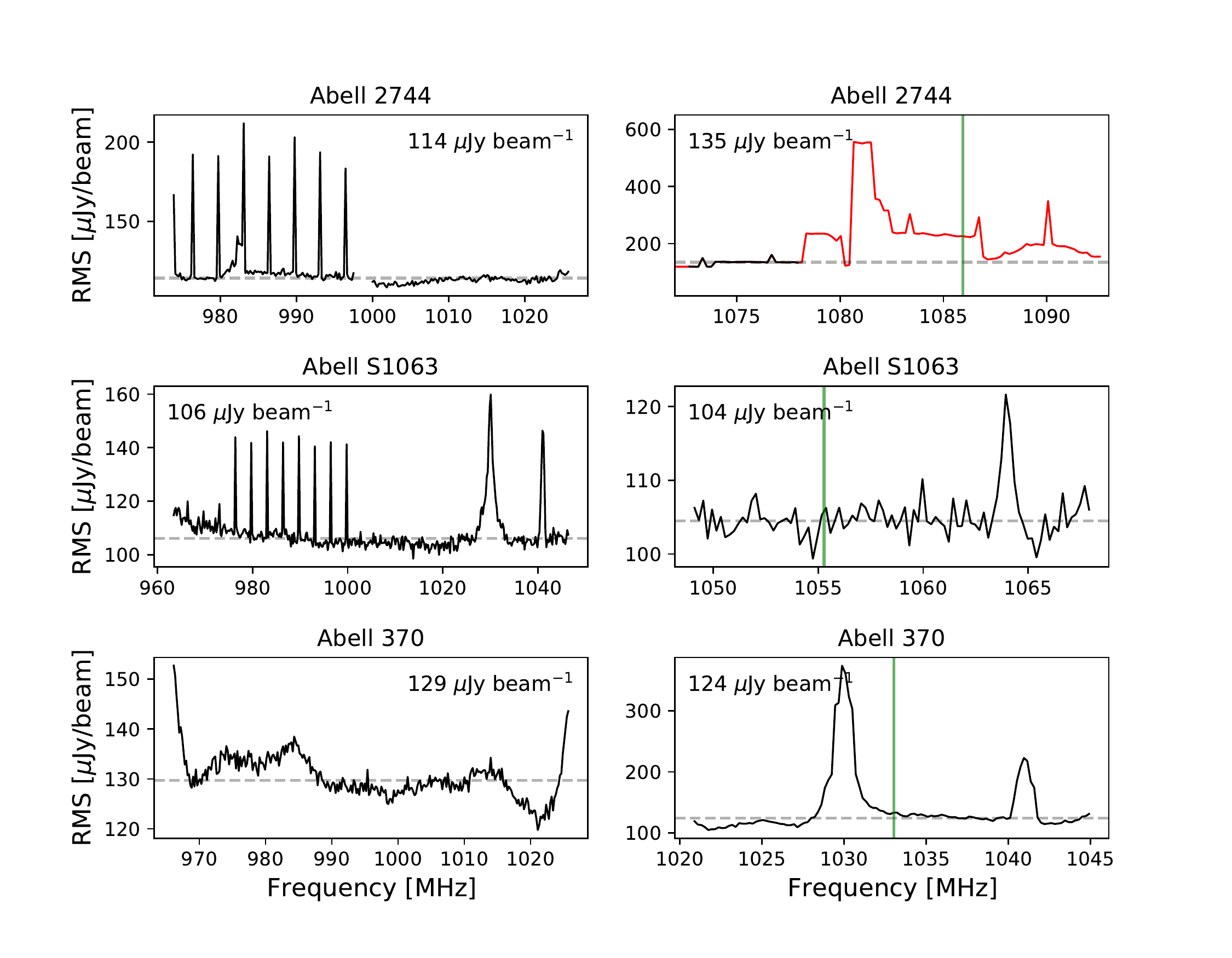}
    \caption{The channel rms for the background cubes (left) and cluster cubes (right) for each cluster. The median rms is indicated on the plot and shown by the dashed horizontal line. The green vertical line shows the frequency corresponding to the cluster redshift. The red region for the Abell~2744 cluster cube indicates the frequency range that was excluded.}
    \label{fig:cuberms}
\end{figure}

\noindent Although the targeted frequency ranges corresponding to the cluster redshifts are mostly clear of RFI, there is particularly bad RFI contamination in the \hi cube of Abell~2744. Fig.~\ref{fig:cuberms} shows the cube rms as a function of frequency, demonstrating that the majority of the cluster redshift range is contaminated. Because of this, we create a subcube of only 27 channels for this cluster (see Table~\ref{tab:4-cubeparameters}).

In Fig.~\ref{fig:cuberms}, there are noticeable periodic single-channel spikes in the cube rms, throughout the majority of the bandwidth of the Abell~2744 and Abell~S1063 cubes. The spikes occur every 16 channels and are attributed to a firmware issue within the MeerKAT correlators at the time of observation, which was within the science verification stage of MeerKAT. In the cases where this is particularly severe, we have flagged the affected channels. The correlator issue was rectified before the observation of Abell~370 and does therefore not affect that observation.
\subsection{Ancillary Data}\label{sssec:4-ancillary}
\begin{figure*}
    \centering
    \includegraphics[scale=0.6]{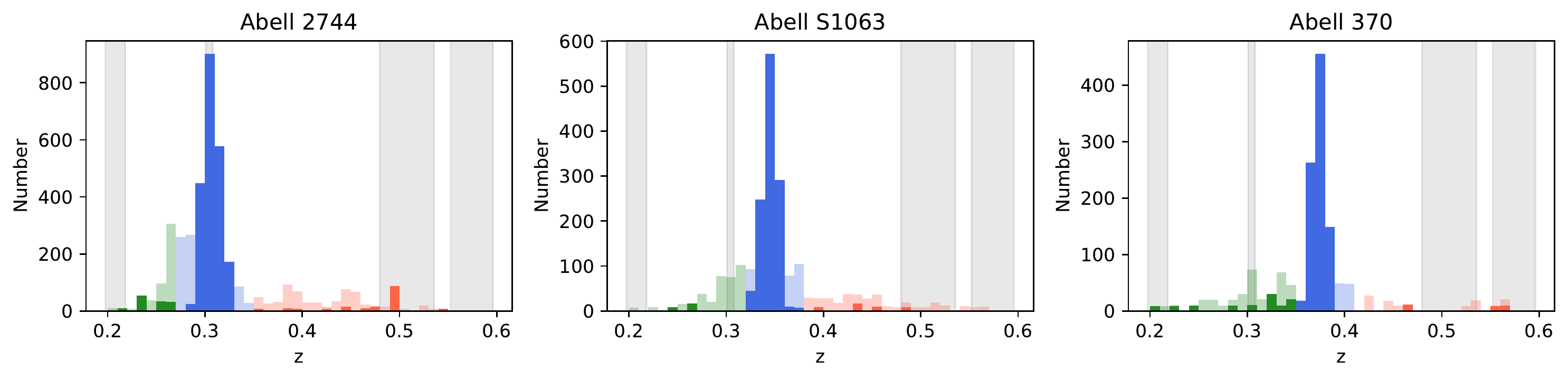}
    \caption[Redshift distribution of objects in the HFF catalogues]{The stellar mass-weighted redshift distribution of objects in the HFF catalogues in the range of $0 < z < 0.58$. The blue represents the galaxies in the cluster redshift range, the green and red represent the objects in the cluster foreground and background respectively, with bin widths of $z=0.02$. The high opacity bars show the sources with spectroscopic redshifts, and the low opacity bars show the objects that only have a photometric redshift. The grey areas indicate frequencies with high RFI occupancy.}
    \label{fig:4-hist}
\end{figure*}

%
The ancillary data used in this work are primarily from The Hubble Frontier Fields Programme catalogues \citep{Shipley_2018}. This catalogue combines data from the \textit{Spitzer} and the \textit{Hubble Space Telescopes}. The optical \textit{HST} observations were obtained using the Advanced Camera for Surveys WFC detector (ACS/WFC), and the near-IR observations were obtained using the Wide Field Camera 3 IR detector (WFC3/IR). These observations were conducted over 7 wide bands, over a wavelength range of $0.35 - 1.7\,\mu$m. Photometric redshifts were derived from the photometry, as described in \citet{Shipley_2018}, and spectroscopic redshifts were compiled from the literature, where available \citep{Owers_2011,Karman_2015,Mahler_2017,Lagattuta_2017}. The photometric and spectroscopic redshift distributions, for the catalogue objects in each cluster, are plotted in Fig.~\ref{fig:4-hist}. The grey areas in Fig.~\ref{fig:4-hist} indicate frequencies with high RFI occupancy due to Global Positioning System (GPS) satellites and Global System for Mobile Communications (GSM) networks. Multiple $<1$ MHz bandwidth intermittent RFI signals from 1000-1200 MHz ($0.18<z<0.42$), not indicated on the plot, may be present due to aircraft transponders.

A sub-catalogue of cluster members was compiled for each cluster. Following \citet{Ma_2008}, cluster membership is assigned to sources that have spectroscopic redshifts in the range of $z = z_\mathrm{cl} \pm 2.5\sigma$, where $z_\mathrm{cl}$ is the mean cluster redshift, and $\sigma$ is the velocity dispersion measured from Fig.~\ref{fig:4-hist}. The number of cluster members for each cluster are listed in Table~\ref{tab:4-cluster}. Another sub-catalogue was created for target lensed sources. Sources were selected if they had a spectroscopic redshift greater than the maximum cluster redshift, as indicated in Fig.~\ref{fig:4-hist}, and less than $z\leq0.5$.
Magnification maps for the HFF clusters have been modelled by multiple independent groups. \citet{Shipley_2018} derive magnification values for each catalogue source from each group's most recent model. In this work, we use the CATS \citep[Clusters As TelescopeS, P.I. Ebeling, e.g.][]{Jauzac_2015} lensing models (Version 4.0), and the magnifications derived therefrom. 

\section{Cluster \hi}\label{sec:4-results}

\subsection{Source Finding}\label{sec:clustersf}
%
Following a visual inspection with no direct \hi detections, we performed a blind search for direct \hi detections in the cluster cubes using \textsc{SoFiA} \citep{Serra_2015}, a software package designed for 3D source-finding in large spectral line cubes. The following \textsc{SoFiA} strategy was used: A low threshold of 2$\sigma$ was selected for the \texttt{Smooth $+$ Clip} algorithm. 3D boxcar smoothing kernels were used, with spatial dimensions of 5, 8 and 10 pixels (i.e. [10, 16, 20] arcsec, which corresponds to [46, 73, 92] kpc at $z=0.35$), and spectral dimensions of 1 and 2 channels. These dimensions were chosen to correspond to the expected velocity width for \hi-massive and/or lensed galaxies, and the PSF FWHM of the cube. Pixels above the detection threshold were assigned to belong to the same source if they had a maximum separation of 2 spatial pixels or 2 frequency channels. 

Using \textsc{SoFiA}'s reliability calculation, and assuming a reliability threshold of 0.95, no reliable detections were identified for any of the clusters. Assuming the median rms values listed in Table~\ref{tab:4-cubeparameters}, we determined the 5$\sigma$ \hi mass detection limit to be $M_\mathrm{HI} = 2.06\times10^{10}\,\mathrm{M}_\odot$ for Abell~2744, $M_\mathrm{HI} = 1.94\times10^{10}\,\mathrm{M}_\odot$ for Abell~S1063 and $M_\mathrm{HI} = 2.31\times10^{10}\,\mathrm{M}_\odot$ for Abell~370. Three candidate galaxies in Abell~S1063 and one in Abell~370 have predicted \hi masses greater than this limit, as estimated from the following equation from \citet{Meyer_2017}:
\begin{equation}\label{eq:intflux}
\centering
   \left(\frac{S}{\mathrm{JyHz}}\right) = \frac{1}{49.7} \left(\frac{M_\mathrm{HI}}{\mathrm{M_\odot}} \right) \left(\frac{D_L}{\mathrm{Mpc}} \right)^{-2},
\end{equation}
and the $M_*-M_\mathrm{HI}$ relation from \citet{Parkash_2018}
\begin{equation}\label{eq:parkash}
    \log M_\mathrm{HI} = 0.51 (\log M_* - 10) + 9.71.
\end{equation}
This relation has a scatter of 0.5 dex, so not all galaxies are guaranteed to be detected.

\subsection{\hi Spectral Stacking}\label{sec:4-stacking}
We use image domain spectral line stacking for sources with spectroscopic redshifts in each cluster. At this angular resolution, sources are not expected to be resolved 
e.g. a mean PSF FWHM of 15 arcsec is equal to 62 kpc at $z=0.3$. The typical predicted diameter for sources in this sample is $\sim 30$ kpc \citep{Wang_2016}, with high mass outliers at $\sim 80$ kpc. Note that sources with redshifts corresponding to fully flagged channels are removed before stacking. The spectra are shifted to rest-frame and a wrapping technique is implemented \citep[e.g.][]{Healy_2019}, such that the original spectrum length is maintained. 
The flux spectra are converted to mass spectra using the following equation \citep[e.g.][]{Delhaize_2013,Hu_2019}:
\begin{equation}\label{eq:conversion}
m_{\mathrm{HI}}(\nu)=4.97 \times 10^{7} S_{\nu} D_{L}^{2} f^{-1}.
\end{equation}
Here, $m_{\mathrm{HI}}$ has units of $\mathrm{M}_\odot \mathrm{MHz}^{-1}$, $S_\nu$ is the rest-frame \hi flux density in Jy, $D_L$ is the luminosity distance in Mpc, and $f$ is the normalised primary beam response. For MeerKAT, the primary beam has an FWHM of $\sim80$' for \hi at $z=0.3$ \citep{Mauch_2020}, and the clusters have a maximum virial radius of $\sim3$' from the centre of the primary beam. Because the entire cluster resides well within the centre of the primary beam, we can approximate the beam pattern as a Gaussian function in all target directions that is constant as a function of frequency.

For stacking, a weight function is introduced, following the procedure used by \citet{Hu_2019}, depending on the rms noise of individual spectra $\sigma$, the primary beam response $f$, and the luminosity distance $D_L$. The weight of the $i^\mathrm{th}$ galaxy is:
\begin{equation}
w_{i}=f^{2} D_{L}^{-\gamma} \sigma^{-2},
\end{equation}
where large values of $\gamma$ give more weight to nearby galaxies and small values of $\gamma$ increase the statistical contribution of more distant galaxies. \citet{Hu_2019} conclude that an optimal stacked SNR is achieved with $\gamma =1$. It should be noted that in general, $\gamma=1$ is not necessarily the best choice for constraining average mass. However, since we cover a small redshift range, and a large SNR is essential, we choose $\gamma=1$. The average stacked mass spectrum is calculated from the following equation:
\begin{equation}\label{Eq:4-avmass}
\left\langle m_{\mathrm{HI}}(v)\right\rangle=\frac{\Sigma_{i=1}^{n} w_{i} m_{\mathrm{HI}, \mathrm{i}}}{\Sigma_{i=1}^{n} w_{i}}.
\end{equation}
The integrated \hi mass $\left\langle M_{\mathrm{HI}}\right\rangle$ is the integral of the spectral peak along the frequency axis:
\begin{equation}\label{inthimass}
M_{\mathrm{HI}}= \int _{\nu_{0}-\Delta \nu}^{\nu_{0}+\Delta \nu}\left\langle m_{\mathrm{HI}}(\nu)\right\rangle d \nu,
\end{equation}
\noindent where $\nu_0$ is the rest frequency of neutral hydrogen. A boxcar function is fit to the peak at rest frequency to estimate $M_{\mathrm{HI}}$ using Bayesian parameter estimation (see Section~\ref{ssec:veracity}). 

Each spectrum is smoothed with a Hanning function in frequency space, with a window of a specified width. Spatial smoothing is also implemented using {\sc Casa's} \texttt{imsmooth}. The smoothing kernel is well-matched to the intrinsic spatial ($1.5\times$PSF FWHM) and velocity (4 channels or $\sim 240$\kms) extent of the targeted sources.

\begin{figure}
    \centering
    \includegraphics[width=0.99\columnwidth]{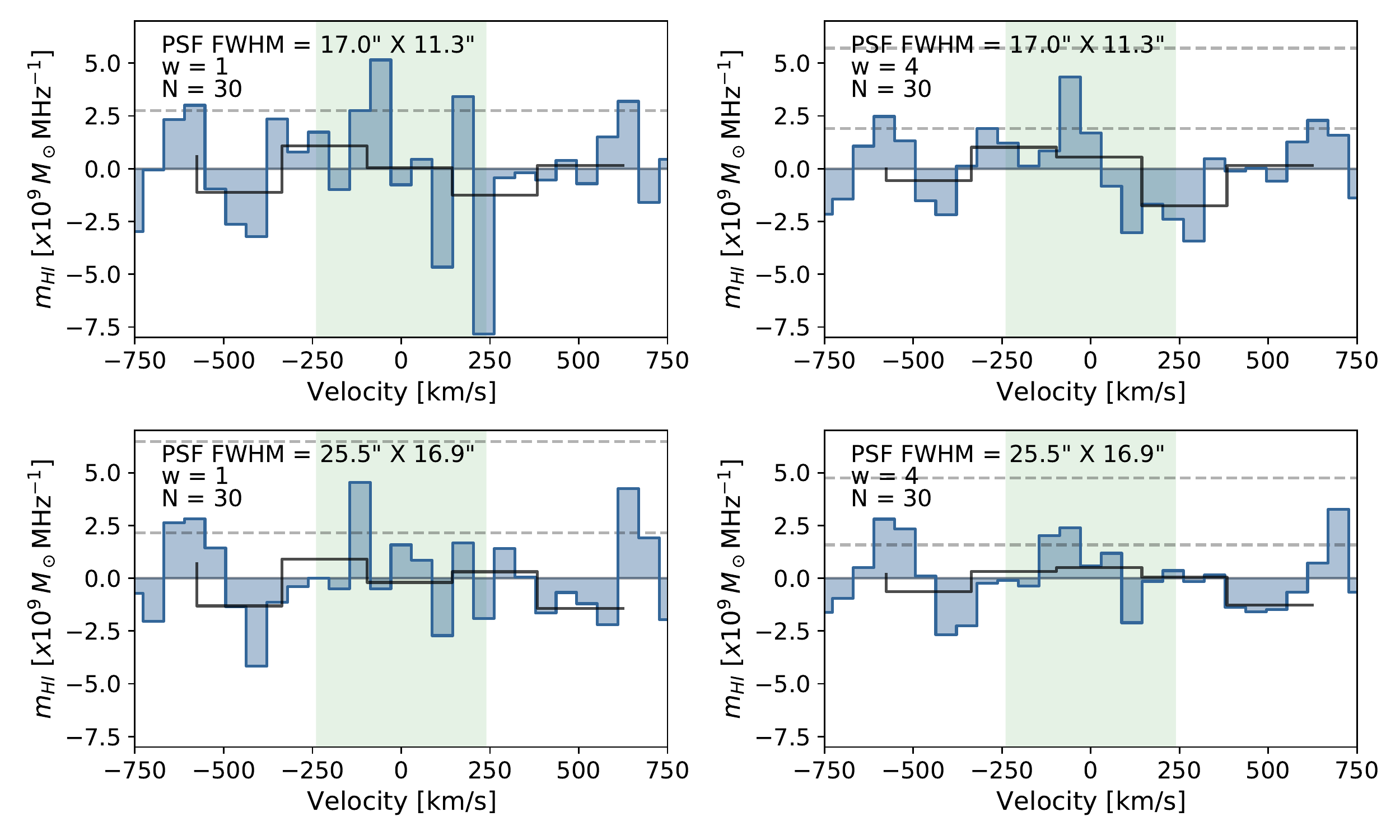}
    \caption[The stacked \hi spectra of galaxies in Abell~2744]{The stacked mass spectra for 30 galaxies with spectroscopic redshifts in Abell~2744. The plots are spatially and spectrally smoothed, as described in Section~\ref{sec:4-stacking}. The smoothed PSF dimensions, spectral smoothing window $w$ in units of channel, and the number of spectra stacked $N$ are shown in the top left corner. The black line shows the spectrum regridded to a velocity resolution of 240~\kms. The green region indicates the velocity range where we expect to detect \hi emission for galaxies of this stellar mass, and the 1$\sigma$ and 3$\sigma$ \hi mass rms are indicated by the grey dashed lines.}
    \label{fig:4-stackresultA2744}
\end{figure}

\begin{figure}
    \centering
    \includegraphics[width=0.99\columnwidth]{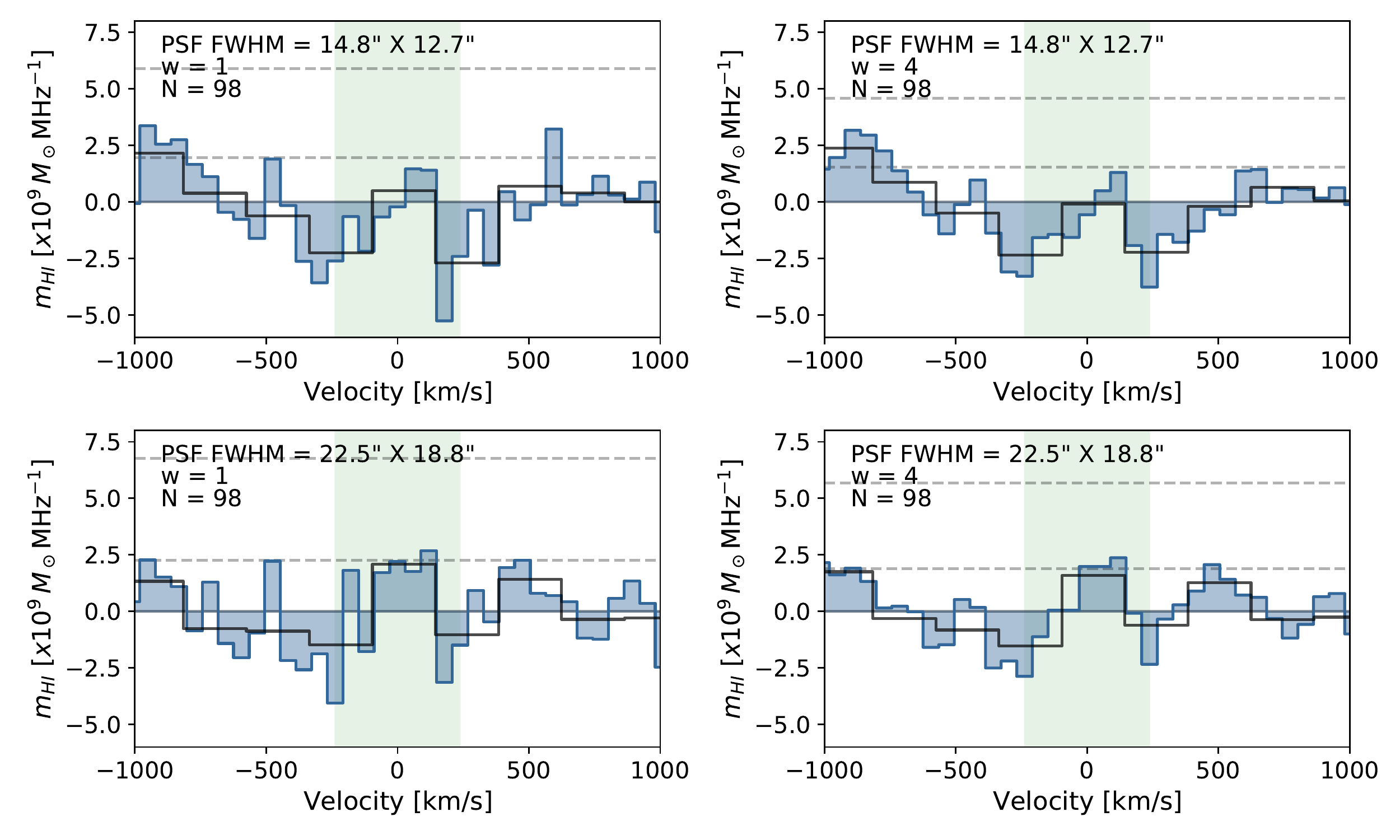}
    \caption[The stacked \hi spectra of galaxies in Abell~S1063]{The stacked mass spectra for 98 galaxies with spectroscopic redshifts in Abell~S1063, as in Fig.~\ref{fig:4-stackresultA2744}.}
    \label{fig:4-stackresultS1063}
\end{figure}

\begin{figure}
    \centering
    \includegraphics[width=0.99\columnwidth]{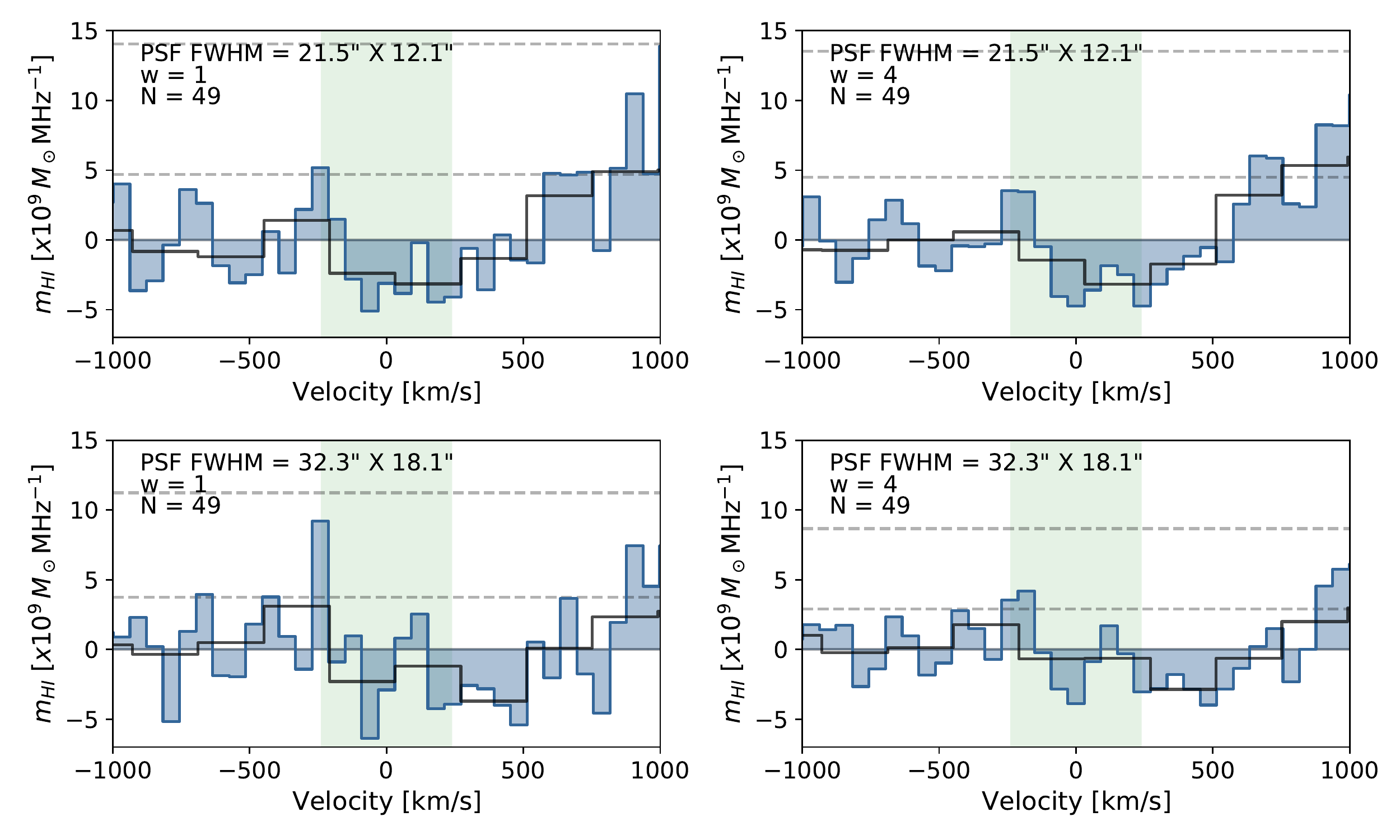}
    \caption[The stacked \hi spectra of galaxies in Abell~370]{The stacked mass spectra for 49 galaxies with spectroscopic redshifts in Abell~370, as in Fig.~\ref{fig:4-stackresultA2744}.}
    \label{fig:4-stackresultA370}
\end{figure}

This stacking procedure was applied to four cubes for each cluster, with varying degrees and types of smoothing:
\begin{enumerate}
    \item unsmoothed,
    \item spatially-smoothed,
    \item spectrally-smoothed, and
    \item spatially- and spectrally-smoothed
\end{enumerate}
We also show the stacked spectra regridded to a velocity resolution of $\sim 240$\kms. The stacking results for all cluster members with spectroscopic redshifts are shown in Figs.~\ref{fig:4-stackresultA2744}, \ref{fig:4-stackresultS1063} and \ref{fig:4-stackresultA370}. The rms of the stacked spectra decreases as approximately $\propto 1/\sqrt{N}$, as expected for Gaussian noise, where $N$ is the number of spectra stacked, for all three cubes.

These spectra show that there are no stacked \hi detections in Abell~2744, Abell~S1063, and Abell~370, for all cluster members with spectroscopic redshifts. For Abell~2744, the rms of the unsmoothed stacked spectrum corresponds to $\sigma = 18\,\mu$Jy \perbeam, and we determine the 5$\sigma$ stacked \hi mass detection limit to be $M_\mathrm{HI} = 2.89\pm 0.21\times10^{9}\,\mathrm{M}_\odot$, well below the knee of the HIMF. For Abell~S1063 and Abell~370, we determine the 5$\sigma$ stacked \hi mass detection limits to be $M_\mathrm{HI} = 2.05\pm 0.04\times10^{9}\,\mathrm{M}_\odot$ ($\sigma = 11\,\mu$Jy \perbeam), and $M_\mathrm{HI} = 4.89\pm 0.19\times10^{9}\,\mathrm{M}_\odot$ ($\sigma = 15\,\mu$Jy \perbeam), respectively. Here, the rms has been converted to flux using Equation~\ref{eq:conversion}, for convenience. The uncertainties are calculated using jackknife resampling with $(N - 1)$ samples, where $N$ is the number of stacked spectra.

\subsubsection{\hi Deficiency}\label{sssec:hidef}

By comparing the predicted \hi mass values for each cluster stack to \hi detection mass limits, we can determine the \hi deficiency parameter $\mathrm{DEF}_\mathrm{HI}$, first introduced by \citet{Haynes_1984}. This parameter is defined as
\begin{equation}\label{hidef}
    \mathrm{DEF}_\mathrm{HI} = \log_{10} \left(\frac{M_\mathrm{HI, exp}}{\mathrm{M}_\odot}\right) - \log_{10} \left(\frac{M_\mathrm{HI, obs}}{\mathrm{M}_\odot}\right),
\end{equation}
where $M_\mathrm{HI, exp}$ and $M_\mathrm{HI, obs}$ are the expected and observed \hi masses, respectively. We predict the expected stacked \hi masses using Equation~\ref{eq:parkash}, assuming a velocity width of 200 \kms and the cube rms levels. While this relation may not hold or be accurate out to these redshifts and in these environments, it does provide a useful comparison. 
We find mean \hi deficiency limits of galaxies in the respective clusters to be $\mathrm{DEF}_\mathrm{HI} > 0.71\pm0.50$ for Abell~2744, $\mathrm{DEF}_\mathrm{HI} > 0.57\pm0.50$ for Abell~S1063, and $\mathrm{DEF}_\mathrm{HI} > 0.33\pm0.50$ for Abell~370. The uncertainties are from the intrinsic scatter of the $M_* - M_\mathrm{HI}$ relation. These values for Abell~2744 and Abell~S1063 fall above $\mathrm{DEF}_\mathrm{HI} > 0.5$. 
Therefore, following the \citet{Cortese_2011} terminology, the galaxies in these clusters can, on average, be classified as \hi deficient with respect to the main sequence of star-formation. The deficiency parameter for Abell~370 falls within the `\hi normal' limits of $-0.5 < \mathrm{DEF}_\mathrm{HI} < 0.5$. However, it should be noted that $\mathrm{DEF}_\mathrm{HI} > 0.33$ is closer to the \hi deficient boundary, and these values are lower limits, with large uncertainties. It is therefore likely that the galaxies in Abell~370 are also, on average, \hi deficient. In addition, we should be mindful of the caveat that Equation~\ref{eq:parkash} is typically applied to samples with a pre-selection of \hi-rich, star-forming galaxies. Deeper observations and further \hi emission detections are required for a better constraint on these values.

\subsubsection{Blue Galaxies}\label{ssec:bluegals}
\begin{figure}
    \centering
    \includegraphics[width=0.99\columnwidth]{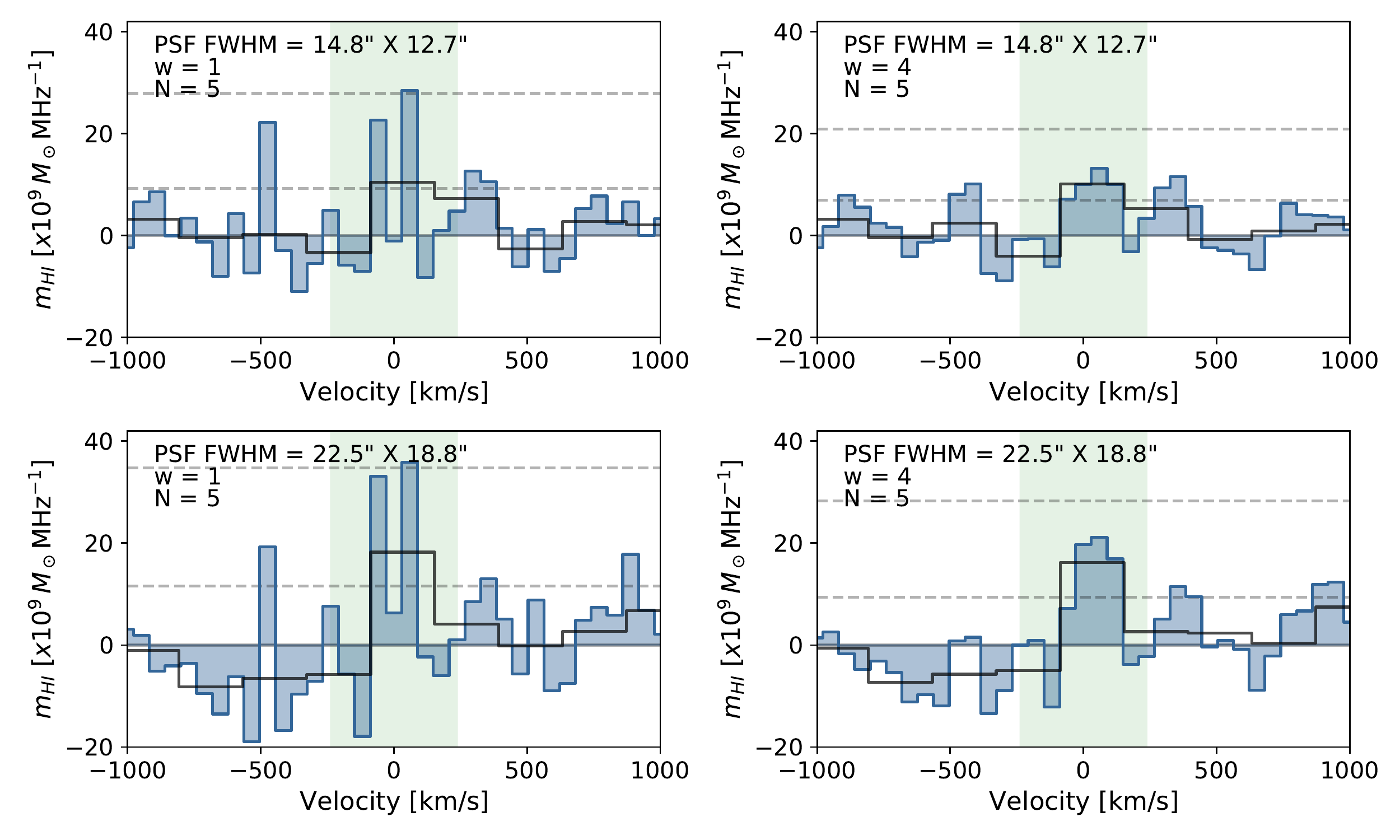}
    \caption[The stacked \hi spectra of blue galaxies in Abell~S1063]{The stacked mass spectra for 5 blue galaxies with spectroscopic redshifts in Abell~S1063, as in Fig.~\ref{fig:4-stackresultA2744}. The grey dashed lines show the 1 and 3$\sigma$ rms.}
    \label{fig:4-bluestackresultAS1063}
\end{figure}

Optically blue, star-forming galaxies typically have higher \hi fractions than red, quiescent galaxies \citep[e.g.][]{Cortese_2011,Brown_2015}. It has been found that stacking only blue galaxies results in higher SNR stacked detections when compared to stacking the full sample \citep[e.g.][]{Kanekar_2016,Bera_2019}, which is indicative of higher gas content in this sub-population. We determine the number of blue galaxies in each cluster sample using the colour-magnitude diagram, classifying sources as blue for $(b~-~v)~<~1$, based on the separation from the red sequence for this statistically small sample. While this is not consistent with typical classifications in other analyses, it has achieved our objective in selecting the bluest of the sample.
Of the cluster members with spectroscopic redshifts, we find the following: Abell~2744 has 7 blue galaxies, and Abell~370 has 3 blue galaxies, all of which do not fall in the redshift range of the limited subcube or are at fully flagged frequencies. Abell~S1063 has 5 blue galaxies, all of which fall into the cluster cube frequency range. The stacked spectrum of these galaxies is shown in Fig.~\ref{fig:4-bluestackresultAS1063}.

We make what appears to be a 3$\sigma$ stacked detection for blue galaxies in Abell~S1063. This peak is visible at all shown degrees of smoothing but is significantly boosted for the spatially smoothed cube. We explore the purported detection in the following sections. 

\subsubsection{Proximity of Sources}
\begin{figure}
    \centering
    \includegraphics[width=0.95\columnwidth]{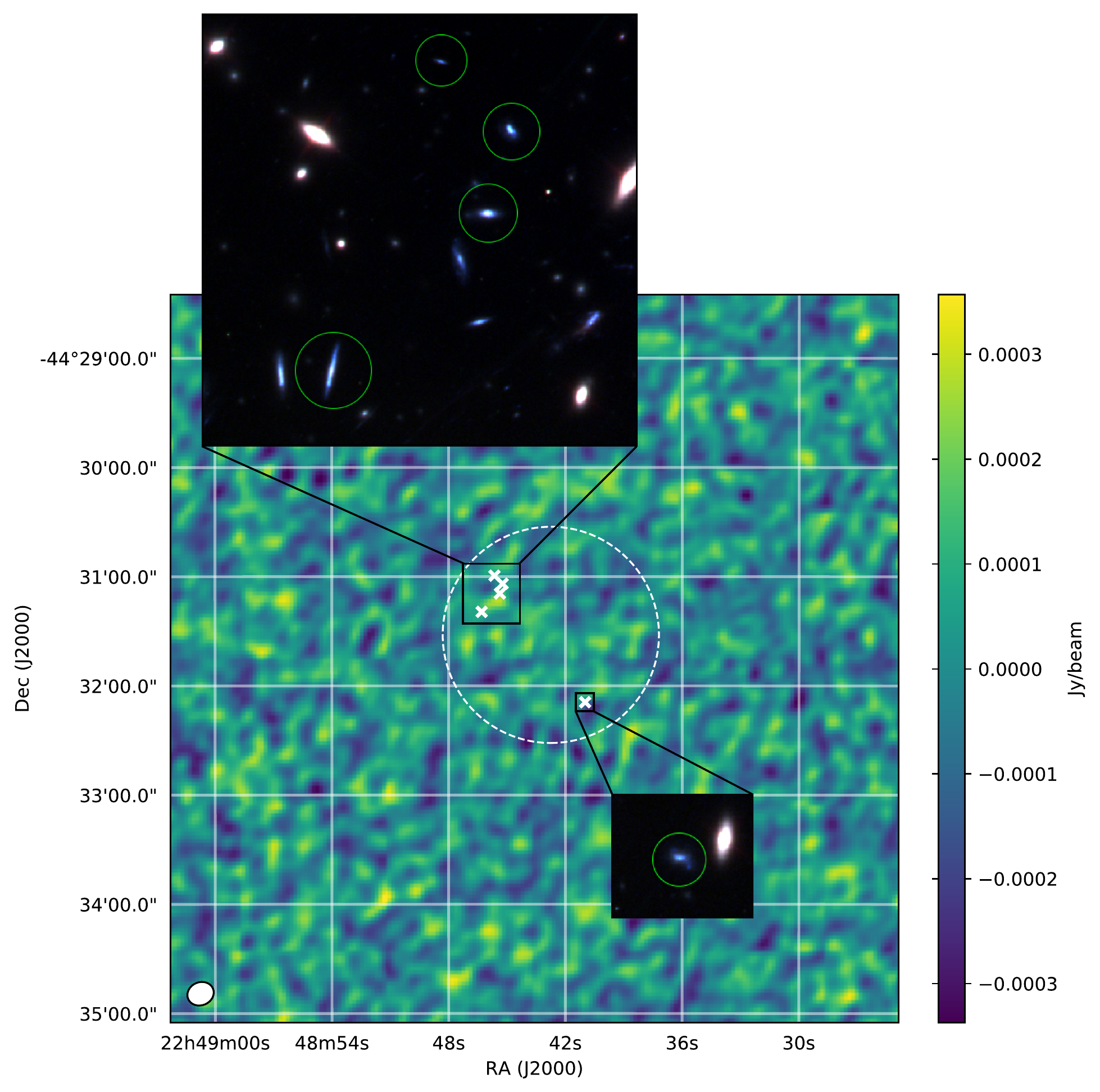}%
    \caption[Coordinate positions of the stacked blue galaxies in Abell~S1063]{MeerKAT channel map of Abell~S1063 corresponding to $z = 0.3421$. The white crosses indicate the positions of the 5 blue galaxies, and the dashed ellipse shows the approximate virial radius of the cluster. The restoring beam is shown in the bottom left. The insets show the RGB $HST$ images of the indicated regions. The stacked sources are indicated by the green circles.}
    \label{fig:4-blueonphot}
\end{figure}

\begin{figure}
    \centering
    \includegraphics[width=0.89\columnwidth]{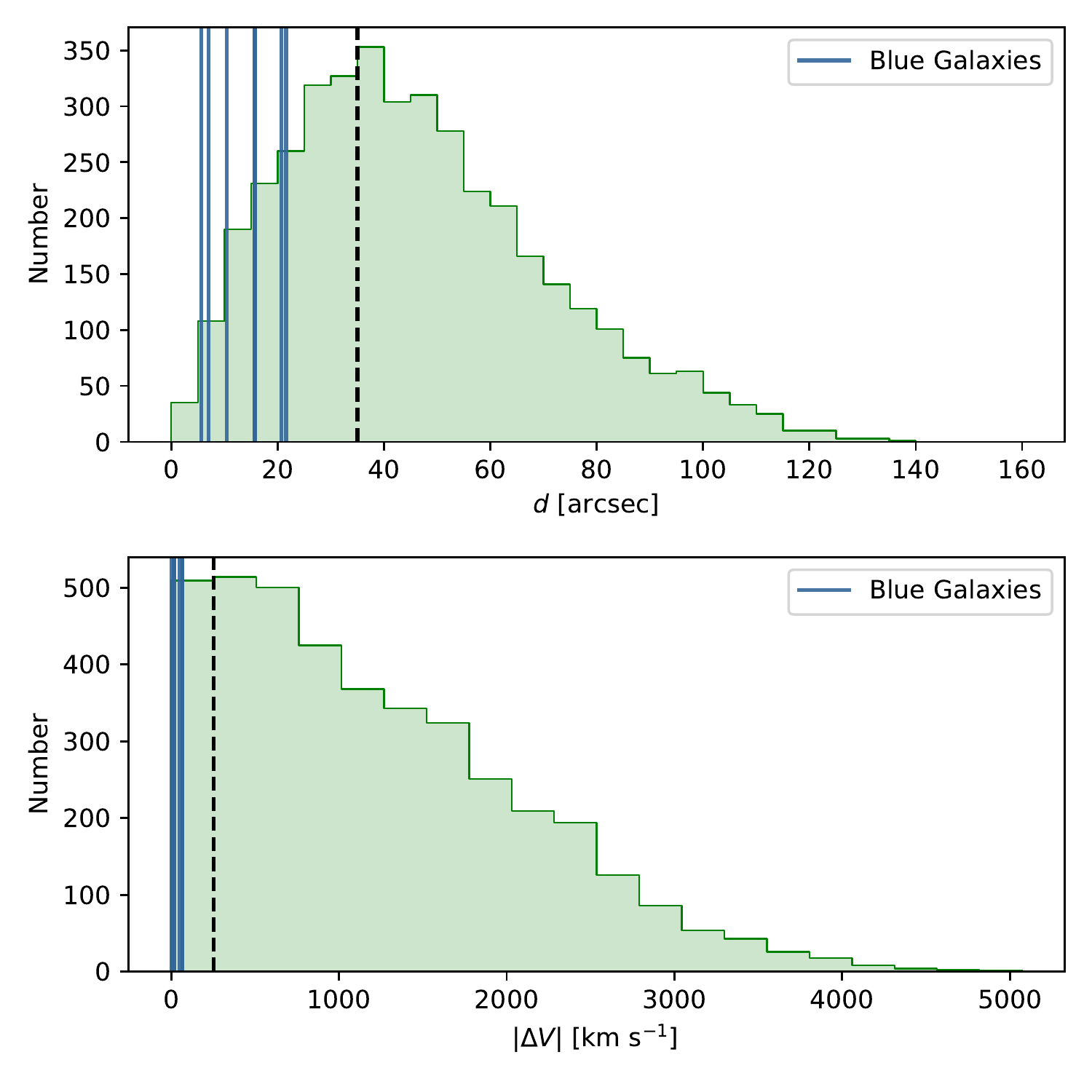}
    \caption[Galaxy separation in Abell~S1063]{Histogram distribution of the projected spatial (top) and recession velocity (bottom) separation between all HFF galaxies in Abell~S1063 (green). The separations between the four nearby blue galaxies are shown in blue. The dashed lines indicate the modal spatial separation and velocity.}
    \label{fig:4-separation}
\end{figure}

Four of the five blue galaxies in Abell~S1063 are located within 30" (corresponding to 135 kpc in projection at $z = 0.34$), and 67 \kms of one another. The coordinates of these sources are over-plotted on the MeerKAT channel map corresponding to the median redshift of $z = 0.3421$ (Fig.~\ref{fig:4-blueonphot}). To place this proximity into perspective with the cluster members as a whole, we plot the distribution of both the spatial and recession velocity separations between all galaxy pairs in the sample. We indicate the separation between the four nearby blue galaxies (i.e. the separation between six pairs of galaxies) on the full sample distributions in Fig.~\ref{fig:4-separation}. As expected, we find that the spatial and velocity separation between the blue galaxies is below the modal separation of the sample. This, and the fact that four of the blue galaxies are heavily clustered near the virial radius of Abell~S1063 strongly suggests that this is a recently in-fallen group \citep[e.g.][]{Jaffe_2015}. Despite the proximity of the four sources, we do not observe any obvious tidal features in the HFF images that would suggest ongoing merger activity. 

\subsubsection{Veracity of Stacked Detection}\label{ssec:veracity}
\begin{figure}
    \centering
    \includegraphics[width=0.89\columnwidth]{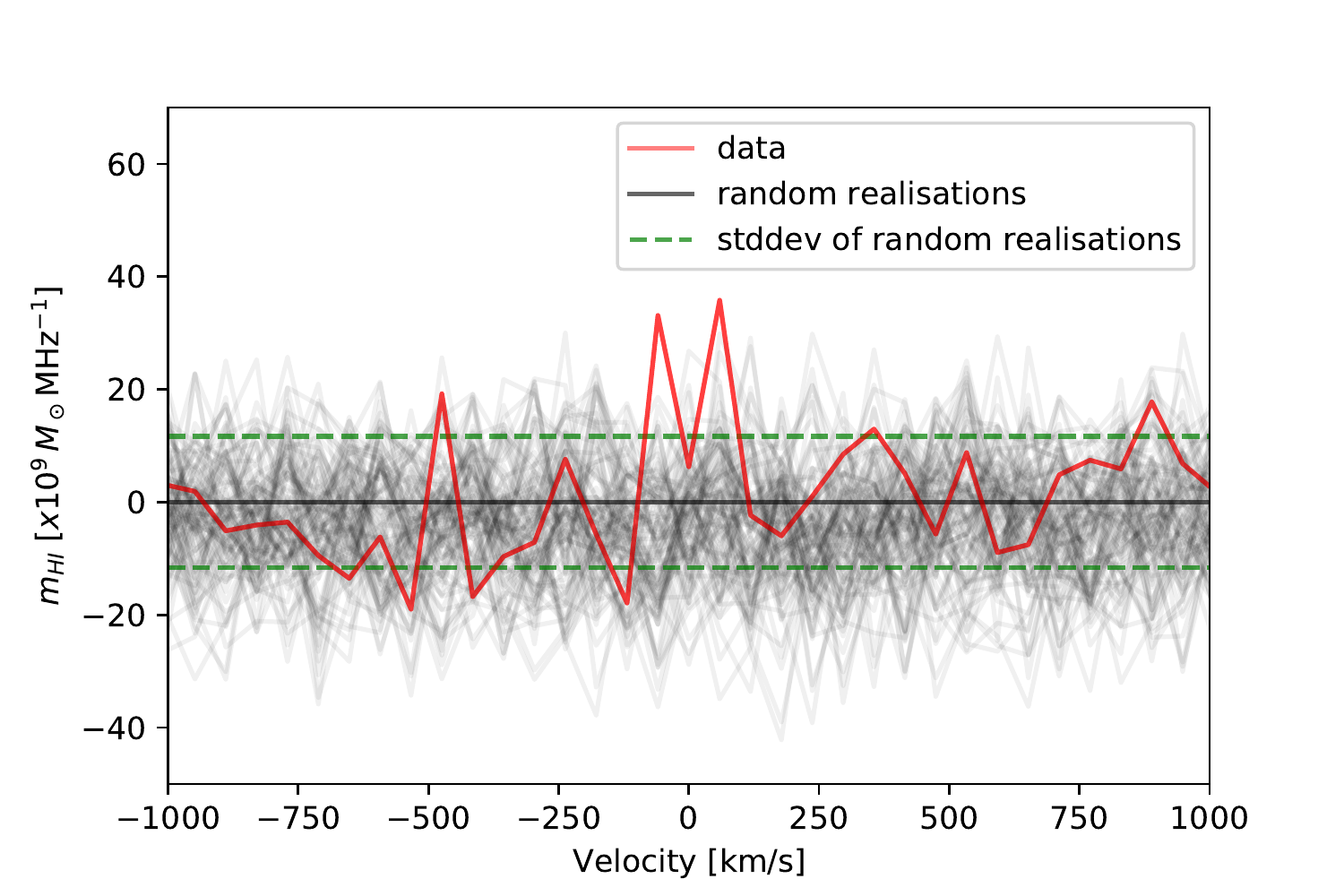}
    \caption[Veracity test for stacked detection in Abell~S1063]{100 random realisations of 5 stacked spectra (black), centred at $z = 0.3421$ in Abell~S1063. The positions of the five extracted spectra for each randomisation follows the same separation of the blue galaxies as shown in Fig.~\ref{fig:4-blueonphot}. The spectra have been extracted from a spatially smoothed cube, and the stacked spectrum for the blue galaxies is shown in red (as the in bottom left in Fig.~\ref{fig:4-bluestackresultAS1063}). The green dashed line indicates the standard deviation of the random realisations.}
    \label{fig:4-stacktest}
\end{figure}

We test the veracity of this 3$\sigma$ detection to rule out the possibility of the peak being caused by the covariance of local image-plane noise. We do this by stacking five randomly selected spectra centred at $z = 0.3421$, that have the same relative separation of the blue galaxies. These coordinates are drawn from a uniformly random distribution, within the spatial extent of the \hi cube. The result of 100 random sets of stacked spectra is shown in Fig.~\ref{fig:4-stacktest} (grey lines). We find that none of these random realisations generates a stacked signal greater in significance than the data spectrum in the $v = \pm 200$\kms$\,$ range, further strengthening the veracity of the detection. In addition, it is encouraging that the standard deviation of the random realisations is similar to the 1$\sigma$ of the stacked spectrum, showing the Gaussianity of the noise.

\begin{figure}
    \centering
    \includegraphics[width=0.8\columnwidth]{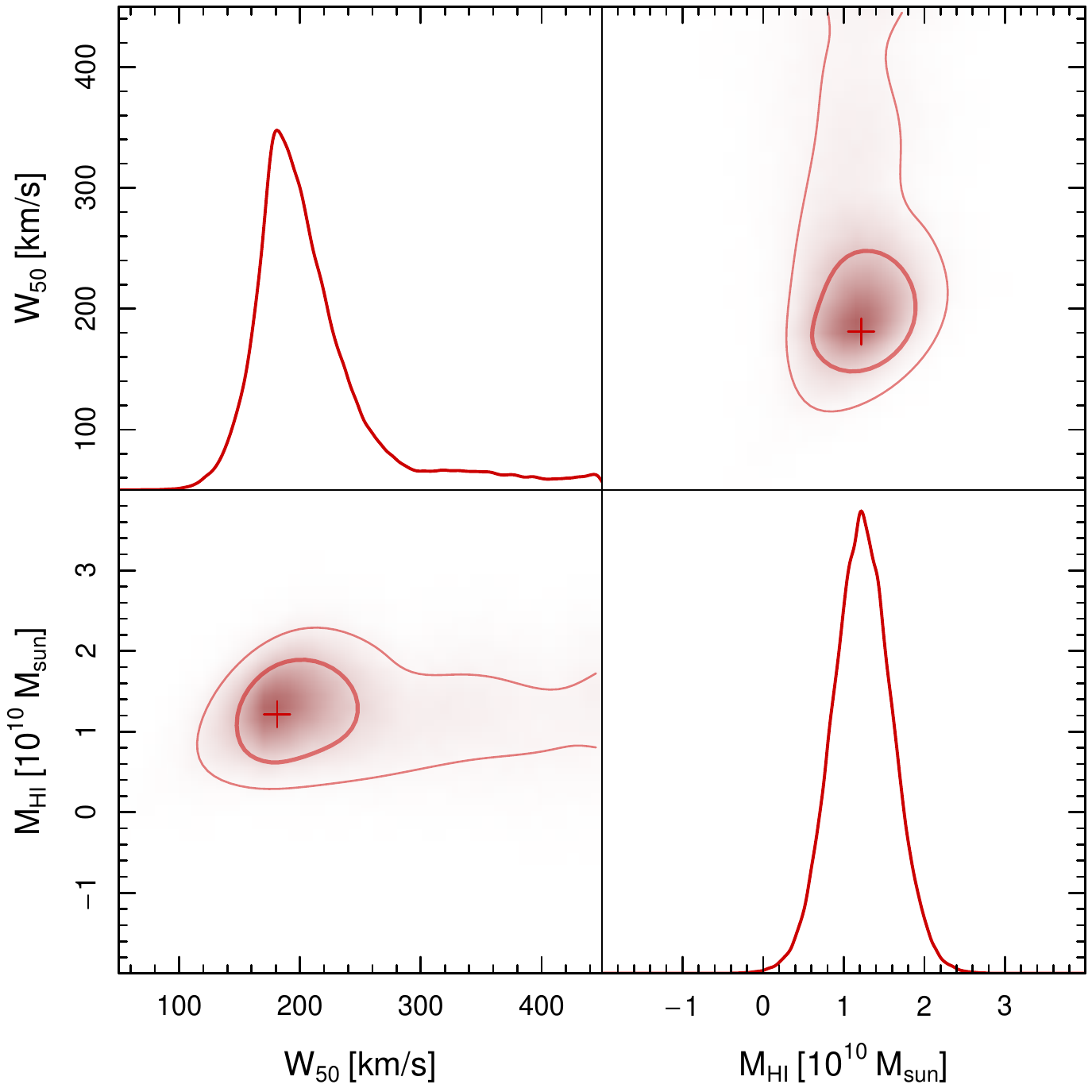}
    \caption{The posteriors for the Bayesian modelling of $M_\mathrm{HI}$ and $w_{50}$. The maximum likelihood solutions are indicated by the crosses. The thick and thin contours respectively show the 68 per cent and 95 per cent posterior probabilities.}
    \label{fig:4-mcmc}
\end{figure}

As a further test of the robustness of the apparent stacked detection, we utilise a Bayesian framework for parameter estimation. To do so, we use a boxcar function whose width and height are expressed in terms of $w_{50}$ and $M_\mathrm{HI}$. This function has the favourable property of retaining the velocity for Gaussian and double-horned intrinsic spectra. The posteriors of these two model parameters are sampled using a Markov Chain Monte Carlo (MCMC) algorithm, with uniform priors. Assuming this boxcar model, we find the maximum likelihood estimates to be $M_\mathrm{HI} = 1.22^{+0.38}_{-0.36}\times 10^{10}\,\mathrm{M}_\odot$ and $w_{50} = 181^{+68}_{-11}$\,\kms, where the sub- and super-script values denote the 16$^{th}$ and 84$^{th}$ percentiles of the posterior distributions shown in Fig.~\ref{fig:4-mcmc}. These distributions are consistent with the estimated 3$\sigma$ significance of the possible detection. This value is significantly larger than the predicted mean \hi mass of the five sources, $M_\mathrm{HI} = 0.10^{+0.32}_{-0.03}\times10 ^{10}\,\mathrm{M}_\odot$ \citep{Parkash_2018}.

In addition to the above parameter estimation of a boxcar model, we also use Bayesian Model Selection to compute the evidence ratio between a boxcar function and flat-line model (i.e. a null test of a constant value). We compute a Bayes factor of 50 between boxcar and flat-line model, which could be considered as `very strong' evidence for the boxcar function hypothesis on the Jeffreys Scale. Assuming the boxcar model to be true, the probability of there being a positive \hi mass is 99.8 per cent. 

Within the limitations of the models employed, the Bayesian approach adds additional quantified rigour to the claimed detection. Deeper observations, preferably at high spatial and velocity resolution will be ultimately required to confirm this. 

\section{Gravitationally-Lensed \hi}\label{sec:gravlens}
In addition to searching for \hi in the clusters, we utilise the strong gravitational lensing properties of these massive clusters to search for magnified distant \hi galaxies behind the clusters. 

\subsection{Predictions and Targeted Search}
\begin{figure*}
    \centering
    \includegraphics[scale=0.55]{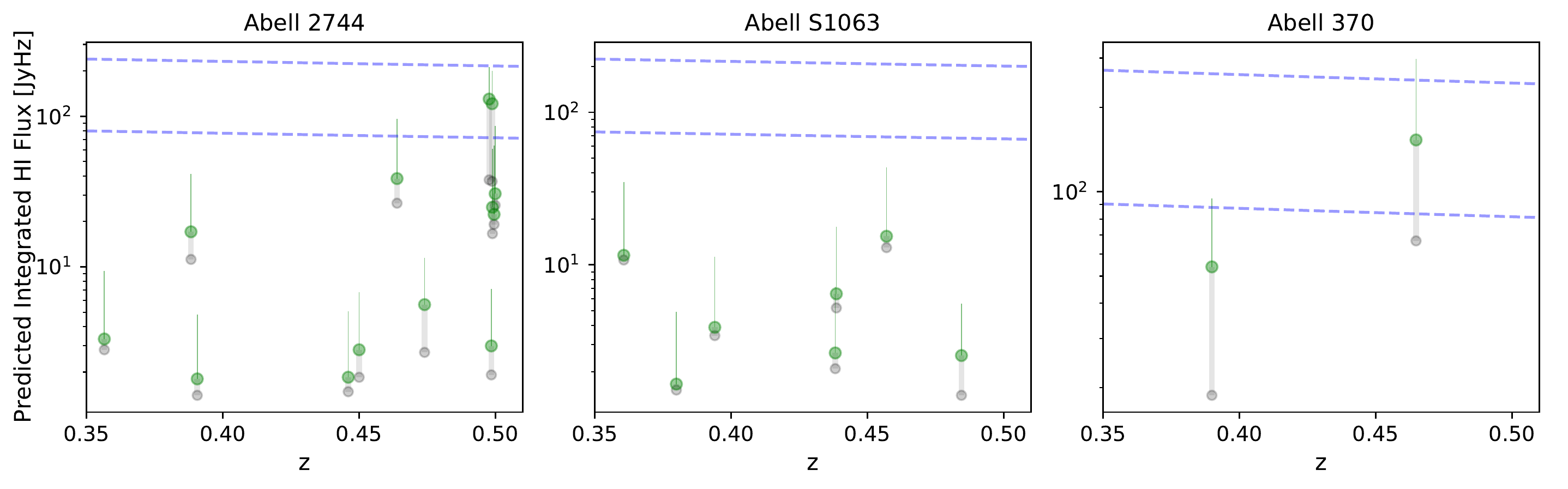}
    \caption[Predicted magnified \hi flux for the objects behind the HFF clusters]{The predicted integrated flux (grey), and magnified integrated flux (green) of the objects lensed by the HFF clusters, as a function of redshift ($z<0.5$). For clarity, the grey bars connect the magnified and unmagnified points of the same source. We assume magnifications from CATS. The blue dashed lines show the 1$\sigma$ and 3$\sigma$ detection limits, and the upper error on the magnified flux are shown in green.}
    \label{fig:5-lenspredict}
\end{figure*}

We modelled the predicted frequency-integrated \hi flux $S_\mathrm{HI}$ [JyHz] of the 22 gravitationally lensed sources with known spectroscopic redshifts between each cluster and $z=0.5$. As before, the integrated flux was estimated using Equation~\ref{eq:parkash}, assuming a velocity width of 200 \kms and the rms of the cubes. Assuming that the observed integrated \hi flux scales linearly with magnification $\mu$ (i.e. all sources are unresolved), we estimated the magnified integrated \hi flux for each of these sources, assuming the CATS \citep{Jauzac_2015} lens model. The predicted results for the three HFF clusters are plotted in Fig.~\ref{fig:5-lenspredict}. From the predictions, we expect no detections greater than 3$\sigma$ for all three clusters. Only 12 of the 22 targets were investigated due to the limits of the cube frequency ranges and RFI contamination. Based on the coordinates and redshifts of targets with known spectroscopic redshifts, we detected no lensed \hi emission for any of the clusters. This is consistent with the models in Blecher et al. (in prep.), who predict larger magnified \hi masses for certain higher redshift targets beyond the frequency range of these L-band observations, the majority of which are redshifted into the UHF-band.
\subsection{Blind Search}
We also performed a blind search for direct \hi detection for lensed, but OIR faint, yet \hi-massive objects, in the redshift range behind the clusters. We use the {\sc SoFiA} software with the \texttt{Smooth + Clip} algorithm, as in Section~\ref{sec:clustersf}, on the `background' cubes. Continuum artefact filtering was also implemented. 
The source finding covered cosmological volumes of 28 815 Mpc$^3$ for Abell~2744, 44 775 Mpc$^3$ for Abell~S1063, and 31 713 Mpc$^3$ for Abell~370.
For all three clusters, no significant detections were found. Using the median cube rms (Table~\ref{tab:4-cubeparameters}), we determine the 5$\sigma$ \hi mass detection limits at $z=0.45$ to be $M_\mathrm{HI} = 3.89\times10^{10}\,\mu^{-1}\mathrm{M}_\odot$, $M_\mathrm{HI} = 3.78\times10^{10}\,\mu^{-1}\mathrm{M}_\odot$ and $M_\mathrm{HI} = 4.43\times10^{10}\,\mu^{-1}\mathrm{M}_\odot$ for Abell~2744, Abell~S1063 and Abell~370.
\section{Discussion}\label{sec:4-discussion}

After stacking the galaxies with spectroscopic redshifts in each cluster (Section~\ref{sec:4-stacking}), we then stacked only the blue galaxies in Abell~S1063 (Section~\ref{ssec:bluegals}) and found the average \hi mass of blue galaxies in Abell~S1063 to be significantly larger than expected. \citet{Elson_2019} show that stacked \hi masses are found to be overestimated in comparison to true \hi results. Due to the low angular resolution \hi cube in comparison with the angular extent of galaxies at high redshifts, the accuracy of stacked \hi measurements can be greatly affected by source confusion \citep[e.g.][]{Delhaize_2013}. Using simulated \hi data cubes, it is shown in \citet{Elson_2016} that the overestimation of co-added \hi mass, caused by source confusion, is more prevalent at lower angular resolution, as one would expect. Smoothing the cubes to poorer angular resolution has been shown to artificially boost the stacked \hi signal (e.g. Fig.~\ref{fig:4-bluestackresultAS1063}), but can simultaneously decrease the accuracy of the stacked \hi mass. The small spatial (138 kpc) separation between the stacked sources (Fig.~\ref{fig:4-blueonphot}) has likely resulted in source confusion. In addition to this, as shown in Fig.~\ref{fig:4-blueonphot}, there are other visibly blue galaxies in the close vicinity of the stacked blue galaxies. These are excluded due to unknown spectroscopic redshifts, however their photometric redshifts suggest that they are cluster members and their \hi flux could be contributing to the confusion. We do not see the characteristic emission peak broadening of this effect in Fig.~\ref{fig:4-bluestackresultAS1063} at our low SNR, as there are few stacked sources. Due to this, the measurement of $M_\mathrm{HI} = 1.22^{+0.38}_{-0.36}\times 10 ^{10}\,\mathrm{M}_\odot$ cannot be taken as an accurate representation of the \hi mass in Abell~S1063. A detailed investigation and modelling of source confusion in this stacked measurement will be explored in future work with deeper observations and 32k correlator mode.

This is one of the few studies of the \hi content of intermediate redshift galaxy clusters, which makes it difficult to make direct comparisons to the literature. However, we can compare to the stacked \hi detection in Abell~370 by \citet{Lah_2009}. They make a 2.7$\sigma$ detection amounting to an average mass of $M_\mathrm{HI} = 4.8 \pm 1.8\times 10 ^{9}\, \mathrm{M}_\odot$, for data smoothed such that all sources are unresolved, spectra binned to $\Delta V = 600$\,\kms, and a sample consisting of 324 galaxies. This measurement is consistent with our upper \hi mass limit for Abell~370. However, the sample of sources used in \citet{Lah_2009} extends beyond the cluster into the area of the GMRT primary beam at 10\% (58.5'), significantly further than the public HFF sample (within 2' of the cluster centre). A more appropriate comparison would be to the \citet{Lah_2009} stack of red galaxies since 48 of our 49 candidate sources are classified as red, following the \citet{Lah_2009} classification of $(b - v ) > 0.57$. Here, they make what they claim to be a marginal 1.2$\sigma$ detection of $M_\mathrm{HI} = 2.6 \pm 2.1\times 10 ^{9}\, \mathrm{M}_\odot$, which is below the sensitivity limits of this experiment.

In this work, we assume the $M_*-M_\mathrm{HI}$ relation from \citet{Parkash_2018}, which is based on low-redshift, star-forming field galaxies, whereas our sample includes only intermediate-redshift cluster galaxies. This is reflected in the \hi deficiencies calculated in Section~\ref{sssec:hidef}. Further deep observations of \hi in intermediate-redshift clusters will test the  appropriateness of this relation and others \citep[e.g.][]{Catinella_2010,Maddox_2014}, as well as their dependability and evolution with redshift and environment. Since galaxy clusters are dynamic environments, they host various processes that remove cold gas from galaxies. This is particularly prevalent in clusters with high X-ray luminosities, as found in Abell~2744, with $L_\mathrm{X} = 3.1 \times 10^{38}\,\mathrm{W}$ \citep{Allen_1998}. 
A modification to the $M_* - M_\mathrm{HI}$ relation would be needed to account for the \hi deficiency in cluster galaxies to accurately predict their \hi masses. Nonetheless, the mass limits probed and the stacked detection shows that this will now be possible at intermediate redshifts.

For all three clusters, there is an inconsistency between the observed rms values and the sensitivity limits obtained from the MeerKAT sensitivity calculator\footnote{\url{https://skaafrica.atlassian.net/servicedesk/customer/portals}}, 
with the observed rms values $\sim55\%$ greater than the predicted values. A major contribution to this inconsistency is likely the bright and extended radio sources associated with these galaxy clusters, as well as an unfavourable Declination for Abell~370. Another contribution is the loss of data due to RFI, particularly for Abell~2744, and the MeerKAT correlator firmware issue. Alternatively, we can investigate the limitations of the calibration strategies used. All cluster measurement sets have $\sim20\%$ flagged visibilities in the considered frequency range of this work. This increases the rms noise by $\sim12\%$, for continuum images. The increase of rms can also be attributed to the propagation of errors in the gain amplitude and flux scaling calibrations. The addition of direction-dependent calibration is expected to further improve the rms noise of the cubes, by mitigating the effect of the primary beam rotation, pointing errors and ionospheric issues. This will be explored in future works and deeper integrations on the HFF.


\section{Conclusion}\label{sec:4-conclusion}
In this paper, we aimed to detect \hi in Abell~2744, Abell~S1063 and Abell~370, all massive galaxies clusters at $0.3<z<0.5$, using data from the MeerKAT Galaxy Cluster Legacy Survey. Through this, we aimed to achieve a better understanding of the \hi content of galaxies in clusters at intermediate redshifts. 
No direct \hi detections were made in the MeerKAT \hi cluster cubes, and we determined 3$\sigma$ \hi mass detection limits down to the knee of the HIMF.

We stacked the spectra of all sources with spectroscopic redshifts in the cube frequency ranges and found that the rms decreases approximately as expected, but detected no stacked emission in the central regions of all three clusters. We determined the stacked 5$\sigma$ \hi mass limits to be $M_\mathrm{HI} = 2.89\pm 0.21\times10^{9}\,\mathrm{M}_\odot$, 
$M_\mathrm{HI} = 2.05\pm 0.04\times10^{9}\,\mathrm{M}_\odot$, and
$M_\mathrm{HI} = 4.89\pm 0.19\times10^{9}\,\mathrm{M}_\odot$ for Abell~2744, Abell~S1063 and Abell~370, respectively. We stacked the spectra of the five blue galaxies with spectroscopic redshifts in Abell~S1063, and made a marginal 3$\sigma$ detection, at a velocity resolution of 60\kms. We determined the average stacked \hi mass to be $M_\mathrm{HI} = 1.22^{+0.38}_{-0.36}\,\times 10^{10}\,\mathrm{M}_\odot$. This is larger,  by a factor of 10, than the predicted \hi masses for the sources from the \citet{Parkash_2018} relation, assuming this $M_*-M_\mathrm{HI}$ relation holds for galaxy clusters. Because of the large uncertainties for the predicted \hi masses, we found that this large stacked \hi mass is an overestimation, attributed to source confusion and possible environmental effects.

We also performed a search for lensed \hi behind the HFF clusters Abell~2744, Abell~S1063, and Abell~370, in the redshift range of $z_\mathrm{cluster} < z < 0.5$, using data from the MeerKAT Galaxy Cluster Legacy Survey and make no direct detections. 
Despite the lack of lensed \hi detections from these observations, the sensitivity of MeerKAT is sufficient to detect these lensed sources with  a factor of 2 longer integration times. The MeerKAT UHF-band covers the redshift range of $>50$ additional lensed sources behind each cluster, which will be investigated with future observations.

These results demonstrate MeerKAT's capability of probing the intermediate redshift range, for high mass individual detections and statistical $M_\mathrm{HI}$ detections. MeerKAT is able to achieve these mass limits within early science observations, despite relatively short integration times for intermediate redshift observations. While not making direct detections, we have obtained mass limits that paint a highly promising picture for future MeerKAT observations, with longer integration times, of large samples of intermediate redshift galaxy clusters.

\section*{Acknowledgements}

The research of SR and RPD is supported by the South African Research Chairs Initiative (grant ID 77948) of the Department of Science and Innovation and National Research Foundation. SR and RPD acknowledge the financial assistance of the South African Radio Astronomy Observatory (SARAO) towards this research (\url{www.ska.ac.za}). IH acknowledges support from the UK Science and Technology Facilities Council [ST/N000919/1], and from the South African Radio Astronomy Observatory which is a facility of the National Research Foundation (NRF), an agency of the Department of Science and Innovation. DO is a recipient of an Australian Research Council Future Fellowship (FT190100083) funded by the Australian Government. The MeerKAT telescope is operated by the South African Radio Astronomy Observatory, which is a facility of the National Research Foundation, an agency of the Department of Science and Innovation. We acknowledge use of the Inter-University Institute for Data Intensive Astronomy (IDIA) data intensive research cloud for data processing. IDIA is a South African university partnership involving the University of Cape Town, the University of Pretoria and the University of the Western Cape. The authors acknowledge the Centre for High Performance Computing (CHPC), South Africa, for providing computational resources to this research project. This work has made use of the Cube Analysis and Rendering Tool for Astronomy \citep[CARTA;][]{Comrie_2021}. This work is based on data and catalogue products from HFF-DeepSpace, funded by the National Science Foundation and Space Telescope Science Institute (operated by the Association of Universities for Research in Astronomy, Inc., under NASA contract NAS5-26555).

\section*{Data Availability}

The radio data used in this analysis are publicly available through the SARAO archive at \url{https://archive.sarao.ac.za/} under proposal ID SSV-20180624-FC-01. Data and catalogue products from HFF-DeepSpace \citep{Shipley_2018} are publicly available at \url{http://cosmos.phy.tufts.edu/~danilo/HFF/Download.html}. Calibrated images and spectra may be made available upon reasonable request.



\bibliographystyle{mnras}
\bibliography{example} 

\begin{thebibliography}{}
\makeatletter
\relax
\def\mn@urlcharsother{\let\do\@makeother \do\$\do\&\do\#\do\^\do\_\do\%\do\~}
\def\mn@doi{\begingroup\mn@urlcharsother \@ifnextchar [ {\mn@doi@}
  {\mn@doi@[]}}
\def\mn@doi@[#1]#2{\def\@tempa{#1}\ifx\@tempa\@empty \href
  {http://dx.doi.org/#2} {doi:#2}\else \href {http://dx.doi.org/#2} {#1}\fi
  \endgroup}
\def\mn@eprint#1#2{\mn@eprint@#1:#2::\@nil}
\def\mn@eprint@arXiv#1{\href {http://arxiv.org/abs/#1} {{\tt arXiv:#1}}}
\def\mn@eprint@dblp#1{\href {http://dblp.uni-trier.de/rec/bibtex/#1.xml}
  {dblp:#1}}
\def\mn@eprint@#1:#2:#3:#4\@nil{\def\@tempa {#1}\def\@tempb {#2}\def\@tempc
  {#3}\ifx \@tempc \@empty \let \@tempc \@tempb \let \@tempb \@tempa \fi \ifx
  \@tempb \@empty \def\@tempb {arXiv}\fi \@ifundefined
  {mn@eprint@\@tempb}{\@tempb:\@tempc}{\expandafter \expandafter \csname
  mn@eprint@\@tempb\endcsname \expandafter{\@tempc}}}

\bibitem[\protect\citeauthoryear{Allen}{Allen}{1998}]{Allen_1998}
Allen S.~W.,  1998, \mn@doi [MNRAS] {10.1046/j.1365-8711.1998.01358.x}, 296,
  392–406

\bibitem[\protect\citeauthoryear{Bera, Kanekar, Chengalur  \& Bagla}{Bera
  et~al.}{2019}]{Bera_2019}
Bera A.,  Kanekar N.,  Chengalur J.~N.,   Bagla J.~S.,  2019, \mn@doi [ApJ]
  {10.3847/2041-8213/ab3656}, 882, L7

\bibitem[\protect\citeauthoryear{Bialas, Lisker, Olczak, Spurzem  \&
  Kotulla}{Bialas et~al.}{2015}]{Bialas_2015}
Bialas D.,  Lisker T.,  Olczak C.,  Spurzem R.,   Kotulla R.,  2015, \mn@doi
  [A&A] {10.1051/0004-6361/201425235}, 576, A103

\bibitem[\protect\citeauthoryear{Blecher, Deane, Heywood  \&
  Obreschkow}{Blecher et~al.}{2019}]{Blecher_2019}
Blecher T.,  Deane R.,  Heywood I.,   Obreschkow D.,  2019, \mn@doi [MNRAS]
  {10.1093/mnras/stz224}, 484, 3681–3690

\bibitem[\protect\citeauthoryear{{Blyth} et~al.,}{{Blyth}
  et~al.}{2016}]{Blyth_2016}
{Blyth} S.,  et~al., 2016, in MeerKAT Science: On the Pathway to the SKA. p.~4

\bibitem[\protect\citeauthoryear{{Broeils} \& {Rhee}}{{Broeils} \&
  {Rhee}}{1997}]{Broeils_1997}
{Broeils} A.~H.,  {Rhee} M.~H.,  1997, \aap, \href
  {https://ui.adsabs.harvard.edu/abs/1997A&A...324..877B} {324, 877}

\bibitem[\protect\citeauthoryear{Brown, Catinella, Cortese, Kilborn, Haynes  \&
  Giovanelli}{Brown et~al.}{2015}]{Brown_2015}
Brown T.,  Catinella B.,  Cortese L.,  Kilborn V.,  Haynes M.~P.,   Giovanelli
  R.,  2015, \mn@doi [MNRAS] {10.1093/mnras/stv1311}, 452, 2479–2489

\bibitem[\protect\citeauthoryear{Brown et~al.,}{Brown
  et~al.}{2016}]{Brown_2016}
Brown T.,  et~al., 2016, \mn@doi [MNRAS] {10.1093/mnras/stw2991}, 466,
  1275–1289

\bibitem[\protect\citeauthoryear{Carilli \& Walter}{Carilli \&
  Walter}{2013}]{Carilli_2013}
Carilli C.,  Walter F.,  2013, \mn@doi [\araa]
  {10.1146/annurev-astro-082812-140953}, 51, 105–161

\bibitem[\protect\citeauthoryear{Catinella et~al.,}{Catinella
  et~al.}{2010}]{Catinella_2010}
Catinella B.,  et~al., 2010, \mn@doi [MNRAS]
  {10.1111/j.1365-2966.2009.16180.x}, 403, 683–708

\bibitem[\protect\citeauthoryear{Chowdhury, Kanekar, Chengalur, Sethi  \&
  Dwarakanath}{Chowdhury et~al.}{2020}]{Chowdhury_2020}
Chowdhury A.,  Kanekar N.,  Chengalur J.~N.,  Sethi S.,   Dwarakanath K.~S.,
  2020, \mn@doi [Nature] {10.1038/s41586-020-2794-7}, 586, 369

\bibitem[\protect\citeauthoryear{Comrie et~al.,}{Comrie
  et~al.}{2021}]{Comrie_2021}
Comrie A.,  et~al., 2021, {CARTA: The Cube Analysis and Rendering Tool for
  Astronomy}, \mn@doi{10.5281/zenodo.4905459}, \url
  {https://doi.org/10.5281/zenodo.4905459}

\bibitem[\protect\citeauthoryear{Cortese, Catinella, Boissier, Boselli  \&
  Heinis}{Cortese et~al.}{2011}]{Cortese_2011}
Cortese L.,  Catinella B.,  Boissier S.,  Boselli A.,   Heinis S.,  2011,
  \mn@doi [MNRAS] {10.1111/j.1365-2966.2011.18822.x}, 415, 1797–1806

\bibitem[\protect\citeauthoryear{Deane, Obreschkow  \& Heywood}{Deane
  et~al.}{2015}]{Deane_2015}
Deane R.~P.,  Obreschkow D.,   Heywood I.,  2015, \mn@doi [MNRAS: Letters]
  {10.1093/mnrasl/slv086}, 452, L49–L53

\bibitem[\protect\citeauthoryear{Delhaize, Meyer, Staveley-Smith  \&
  Boyle}{Delhaize et~al.}{2013}]{Delhaize_2013}
Delhaize J.,  Meyer M.~J.,  Staveley-Smith L.,   Boyle B.~J.,  2013, \mn@doi
  [MNRAS] {10.1093/mnras/stt810}, 433, 1398–1410

\bibitem[\protect\citeauthoryear{Deshev, Haines, Hwang, Finoguenov, Taylor,
  Orlitova, Einasto  \& Ziegler}{Deshev et~al.}{2020}]{Deshev_2020}
Deshev B.,  Haines C.,  Hwang H.~S.,  Finoguenov A.,  Taylor R.,  Orlitova I.,
  Einasto M.,   Ziegler B.,  2020, \mn@doi [A&A] {10.1051/0004-6361/202037803},
  638, A126

\bibitem[\protect\citeauthoryear{Dressler et~al.,}{Dressler
  et~al.}{1997}]{Dressler_1997}
Dressler A.,  et~al., 1997, \mn@doi [ApJ] {10.1086/304890}, 490, 577–591

\bibitem[\protect\citeauthoryear{Dressler, Smail, Poggianti, Butcher, Couch,
  Ellis  \& Oemler}{Dressler et~al.}{1999}]{Dressler_1999}
Dressler A.,  Smail I.,  Poggianti B.~M.,  Butcher H.,  Couch W.~J.,  Ellis
  R.~S.,   Oemler Jr. A.,  1999, \mn@doi [ApJ Supplement Series]
  {10.1086/313213}, 122, 51–80

\bibitem[\protect\citeauthoryear{Dénes, Kilborn, Koribalski  \& Wong}{Dénes
  et~al.}{2015}]{D_nes_2015}
Dénes H.,  Kilborn V.~A.,  Koribalski B.~S.,   Wong O.~I.,  2015, \mn@doi
  [MNRAS] {10.1093/mnras/stv2391}, 455, 1294–1308

\bibitem[\protect\citeauthoryear{Elson, Blyth  \& Baker}{Elson
  et~al.}{2016}]{Elson_2016}
Elson E.~C.,  Blyth S.~L.,   Baker A.~J.,  2016, \mn@doi [MNRAS]
  {10.1093/mnras/stw1291}, 460, 4366–4381

\bibitem[\protect\citeauthoryear{Elson, Baker  \& Blyth}{Elson
  et~al.}{2019}]{Elson_2019}
Elson E.~C.,  Baker A.~J.,   Blyth S.~L.,  2019, \mn@doi [MNRAS]
  {10.1093/mnras/stz1178}, 486, 4894–4903

\bibitem[\protect\citeauthoryear{Fernández et~al.,}{Fernández
  et~al.}{2016}]{Fern_ndez_2016}
Fernández X.,  et~al., 2016, \mn@doi [ApJ] {10.3847/2041-8205/824/1/l1}, 824,
  L1

\bibitem[\protect\citeauthoryear{Gavazzi, Consolandi, Gutierrez, Boselli  \&
  Yoshida}{Gavazzi et~al.}{2018}]{Gavazzi_2018}
Gavazzi G.,  Consolandi G.,  Gutierrez M.~L.,  Boselli A.,   Yoshida M.,  2018,
  \mn@doi [A&A] {10.1051/0004-6361/201833427}, 618, A130

\bibitem[\protect\citeauthoryear{Gunn \& Gott}{Gunn \& Gott}{1972}]{Gunn_1972}
Gunn J.~E.,  Gott J.~R.,  1972, \mn@doi [ApJ] {10.1086/151605}, 176, 1

\bibitem[\protect\citeauthoryear{{Haynes} \& {Giovanelli}}{{Haynes} \&
  {Giovanelli}}{1984}]{Haynes_1984}
{Haynes} M.~P.,  {Giovanelli} R.,  1984, \mn@doi [\aj] {10.1086/113573}, \href
  {https://ui.adsabs.harvard.edu/abs/1984AJ.....89..758H} {89, 758}

\bibitem[\protect\citeauthoryear{Healy, Blyth, Elson, van Driel, Butcher,
  Schneider, Lehnert  \& Minchin}{Healy et~al.}{2019}]{Healy_2019}
Healy J.,  Blyth S.-L.,  Elson E.,  van Driel W.,  Butcher Z.,  Schneider S.,
  Lehnert M.~D.,   Minchin R.,  2019, \mn@doi [MNRAS] {10.1093/mnras/stz1555},
  487, 4901–4938

\bibitem[\protect\citeauthoryear{{Healy} et~al.,}{{Healy}
  et~al.}{2021}]{Healy_2021}
{Healy} J.,  et~al., 2021, \mn@doi [\aap] {10.1051/0004-6361/202038738}, \href
  {https://ui.adsabs.harvard.edu/abs/2021A&A...650A..76H} {650, A76}

\bibitem[\protect\citeauthoryear{Hess \& Wilcots}{Hess \&
  Wilcots}{2013}]{Hess_2013}
Hess K.~M.,  Wilcots E.~M.,  2013, \mn@doi [AJ] {10.1088/0004-6256/146/5/124},
  146, 124

\bibitem[\protect\citeauthoryear{{Heywood}}{{Heywood}}{2020}]{Heywood_2020}
{Heywood} I.,  2020, {oxkat: Semi-automated imaging of MeerKAT observations}
  (\mn@eprint {ascl} {2009.003})

\bibitem[\protect\citeauthoryear{Hu et~al.,}{Hu et~al.}{2019}]{Hu_2019}
Hu W.,  et~al., 2019, \mn@doi [MNRAS] {10.1093/mnras/stz2038}, 489, 1619–1632

\bibitem[\protect\citeauthoryear{Hunt, Pisano  \& Edel}{Hunt
  et~al.}{2016}]{Hunt_2016}
Hunt L.~R.,  Pisano D.~J.,   Edel S.,  2016, \mn@doi [AJ]
  {10.3847/0004-6256/152/2/30}, 152, 30

\bibitem[\protect\citeauthoryear{Jaff{\'e}, Poggianti, Verheijen, Deshev  \&
  van Gorkom}{Jaff{\'e} et~al.}{2013}]{Jaffe_2013}
Jaff{\'e} Y.~L.,  Poggianti B.~M.,  Verheijen M. A.~W.,  Deshev B.~Z.,   van
  Gorkom J.~H.,  2013, \mn@doi [MNRAS] {10.1093/mnras/stt250}, 431, 2111–2125

\bibitem[\protect\citeauthoryear{Jaff{\'e}, Smith, Candlish, Poggianti, Sheen
  \& Verheijen}{Jaff{\'e} et~al.}{2015}]{Jaffe_2015}
Jaff{\'e} Y.~L.,  Smith R.,  Candlish G.~N.,  Poggianti B.~M.,  Sheen Y.-K.,
  Verheijen M. A.~W.,  2015, \mn@doi [MNRAS] {10.1093/mnras/stv100}, 448,
  1715–1728

\bibitem[\protect\citeauthoryear{Jarvis et~al.,}{Jarvis
  et~al.}{2017}]{Jarvis_2017}
Jarvis M.~J.,  et~al., 2017, The MeerKAT International GHz Tiered Extragalactic
  Exploration (MIGHTEE) Survey (\mn@eprint {arXiv} {1709.01901})

\bibitem[\protect\citeauthoryear{{Jauzac} et~al.,}{{Jauzac}
  et~al.}{2014}]{Jauzac_2015}
{Jauzac} M.,  et~al., 2014, \mn@doi [\mnras] {10.1093/mnras/stu1355}, \href
  {https://ui.adsabs.harvard.edu/abs/2014MNRAS.443.1549J} {443, 1549}

\bibitem[\protect\citeauthoryear{Joshi, Pillepich, Nelson, Marinacci, Springel,
  Rodriguez-Gomez, Vogelsberger  \& Hernquist}{Joshi et~al.}{2020}]{Joshi_2020}
Joshi G.~D.,  Pillepich A.,  Nelson D.,  Marinacci F.,  Springel V.,
  Rodriguez-Gomez V.,  Vogelsberger M.,   Hernquist L.,  2020, \mn@doi [MNRAS]
  {10.1093/mnras/staa1668}, 496, 2673–2703

\bibitem[\protect\citeauthoryear{{J{\'o}zsa} et~al.,}{{J{\'o}zsa}
  et~al.}{2020}]{Jozsa_2020}
{J{\'o}zsa} G. I.~G.,  et~al., 2020, {CARACal: Containerized Automated Radio
  Astronomy Calibration pipeline} (\mn@eprint {ascl} {2006.014})

\bibitem[\protect\citeauthoryear{Kanekar, Sethi  \& Dwarakanath}{Kanekar
  et~al.}{2016}]{Kanekar_2016}
Kanekar N.,  Sethi S.,   Dwarakanath K.~S.,  2016, \mn@doi [ApJ]
  {10.3847/2041-8205/818/2/l28}, 818, L28

\bibitem[\protect\citeauthoryear{Karman et~al.,}{Karman
  et~al.}{2015}]{Karman_2015}
Karman W.,  et~al., 2015, \mn@doi [A&A] {10.1051/0004-6361/201424962}, 574, A11

\bibitem[\protect\citeauthoryear{{Kenyon}, {Smirnov}, {Grobler}  \&
  {Perkins}}{{Kenyon} et~al.}{2018}]{Kenyon_2018}
{Kenyon} J.~S.,  {Smirnov} O.~M.,  {Grobler} T.~L.,   {Perkins} S.~J.,  2018,
  \mn@doi [\mnras] {10.1093/mnras/sty1221}, \href
  {https://ui.adsabs.harvard.edu/abs/2018MNRAS.478.2399K} {478, 2399}

\bibitem[\protect\citeauthoryear{Kere{\v s}, Katz, Weinberg  \& Dave}{Kere{\v
  s} et~al.}{2005}]{Keres_2005}
Kere{\v s} D.,  Katz N.,  Weinberg D.~H.,   Dave R.,  2005, \mn@doi [MNRAS]
  {10.1111/j.1365-2966.2005.09451.x}, 363, 2–28

\bibitem[\protect\citeauthoryear{Kneib \& Natarajan}{Kneib \&
  Natarajan}{2011}]{Kneib_2011}
Kneib J.-P.,  Natarajan P.,  2011, \mn@doi [A&ARv] {10.1007/s00159-011-0047-3},
  19

\bibitem[\protect\citeauthoryear{Knowles et~al.,}{Knowles
  et~al.}{2021}]{Knowles_2021}
Knowles K.,  et~al., 2021, The MeerKAT Galaxy Cluster Legacy Survey I. Survey
  Overview and Highlights (\mn@eprint {arXiv} {2111.05673})

\bibitem[\protect\citeauthoryear{Lagattuta et~al.,}{Lagattuta
  et~al.}{2017}]{Lagattuta_2017}
Lagattuta D.~J.,  et~al., 2017, \mn@doi [MNRAS] {10.1093/mnras/stx1079}, 469,
  3946–3964

\bibitem[\protect\citeauthoryear{Lah et~al.,}{Lah et~al.}{2009}]{Lah_2009}
Lah P.,  et~al., 2009, \mn@doi [MNRAS] {10.1111/j.1365-2966.2009.15368.x}, 399,
  1447–1470

\bibitem[\protect\citeauthoryear{Lee-Waddell et~al.,}{Lee-Waddell
  et~al.}{2017}]{Lee_Waddell_2017}
Lee-Waddell K.,  et~al., 2017, \mn@doi [MNRAS] {10.1093/mnras/stx2808}, 474,
  1108–1115

\bibitem[\protect\citeauthoryear{Leroy, Walter, Brinks, Bigiel, de Blok, Madore
   \& Thornley}{Leroy et~al.}{2008}]{Leroy_2008}
Leroy A.~K.,  Walter F.,  Brinks E.,  Bigiel F.,  de Blok W. J.~G.,  Madore B.,
    Thornley M.~D.,  2008, \mn@doi [AJ] {10.1088/0004-6256/136/6/2782}, 136,
  2782–2845

\bibitem[\protect\citeauthoryear{Li, Obreschkow, Lagos, Cortese, Welker  \&
  Džudžar}{Li et~al.}{2020}]{Li_2020}
Li J.,  Obreschkow D.,  Lagos C.,  Cortese L.,  Welker C.,   Džudžar R.,
  2020, \mn@doi [MNRAS] {10.1093/mnras/staa514}, 493, 5024–5037

\bibitem[\protect\citeauthoryear{Lotz et~al.,}{Lotz et~al.}{2017}]{Lotz_2017}
Lotz J.~M.,  et~al., 2017, \mn@doi [ApJ] {10.3847/1538-4357/837/1/97}, 837, 97

\bibitem[\protect\citeauthoryear{Ma, Ebeling, Donovan  \& Barrett}{Ma
  et~al.}{2008}]{Ma_2008}
Ma C.,  Ebeling H.,  Donovan D.,   Barrett E.,  2008, \mn@doi [ApJ]
  {10.1086/589991}, 684, 160–176

\bibitem[\protect\citeauthoryear{Maddox, Hess, Obreschkow, Jarvis  \&
  Blyth}{Maddox et~al.}{2014}]{Maddox_2014}
Maddox N.,  Hess K.~M.,  Obreschkow D.,  Jarvis M.~J.,   Blyth S.-L.,  2014,
  \mn@doi [MNRAS] {10.1093/mnras/stu2532}, 447, 1610–1617

\bibitem[\protect\citeauthoryear{Mahler et~al.,}{Mahler
  et~al.}{2017}]{Mahler_2017}
Mahler G.,  et~al., 2017, \mn@doi [MNRAS] {10.1093/mnras/stx1971}, 473,
  663–692

\bibitem[\protect\citeauthoryear{Makhathini}{Makhathini}{2018}]{Makhathini_2018}
Makhathini S.,  2018, PhD thesis, Rhodes University, Drosty Rd, Grahamstown,
  6139, Eastern Cape, South Africa

\bibitem[\protect\citeauthoryear{Mauch et~al.,}{Mauch
  et~al.}{2020}]{Mauch_2020}
Mauch T.,  et~al., 2020, \mn@doi [ApJ] {10.3847/1538-4357/ab5d2d}, 888, 61

\bibitem[\protect\citeauthoryear{{McMullin}, {Waters}, {Schiebel}, {Young}  \&
  {Golap}}{{McMullin} et~al.}{2007}]{McMullin_2007}
{McMullin} J.~P.,  {Waters} B.,  {Schiebel} D.,  {Young} W.,   {Golap} K.,
  2007, in {Shaw} R.~A.,  {Hill} F.,   {Bell} D.~J.,  eds,  Astronomical
  Society of the Pacific Conference Series Vol. 376, Astronomical Data Analysis
  Software and Systems XVI. p.~127

\bibitem[\protect\citeauthoryear{Meyer, Robotham, Obreschkow, Westmeier, Duffy
  \& Staveley-Smith}{Meyer et~al.}{2017}]{Meyer_2017}
Meyer M.,  Robotham A.,  Obreschkow D.,  Westmeier T.,  Duffy A.~R.,
  Staveley-Smith L.,  2017, \mn@doi [Publications of the Astronomical Society
  of Australia] {10.1017/pasa.2017.31}, 34

\bibitem[\protect\citeauthoryear{Moore, Lake  \& Katz}{Moore
  et~al.}{1998}]{Moore_1998}
Moore B.,  Lake G.,   Katz N.,  1998, \mn@doi [ApJ] {10.1086/305264}, 495, 139

\bibitem[\protect\citeauthoryear{Noordam \& Smirnov}{Noordam \&
  Smirnov}{2010}]{Noordam_2010}
Noordam J.~E.,  Smirnov O.~M.,  2010, \mn@doi [A&A]
  {10.1051/0004-6361/201015013}, 524, A61

\bibitem[\protect\citeauthoryear{Obreschkow \& Rawlings}{Obreschkow \&
  Rawlings}{2009}]{Obreschkow_2009}
Obreschkow D.,  Rawlings S.,  2009, \mn@doi [ApJ]
  {10.1088/0004-637x/696/2/l129}, 696, L129–L132

\bibitem[\protect\citeauthoryear{Odekon et~al.,}{Odekon
  et~al.}{2016}]{Odekon_2016}
Odekon M.~C.,  et~al., 2016, \mn@doi [ApJ] {10.3847/0004-637x/824/2/110}, 824,
  110

\bibitem[\protect\citeauthoryear{{Offringa}}{{Offringa}}{2010}]{Offringa_2010}
{Offringa} A.~R.,  2010, {AOFlagger: RFI Software} (\mn@eprint {ascl}
  {1010.017})

\bibitem[\protect\citeauthoryear{{Offringa} et~al.,}{{Offringa}
  et~al.}{2014}]{Offringa_2014}
{Offringa} A.~R.,  et~al., 2014, \mn@doi [\mnras] {10.1093/mnras/stu1368},
  \href {https://ui.adsabs.harvard.edu/abs/2014MNRAS.444..606O} {444, 606}

\bibitem[\protect\citeauthoryear{Oosterloo \& van Gorkom}{Oosterloo \& van
  Gorkom}{2005}]{Oosterloo_2005}
Oosterloo T.,  van Gorkom J.,  2005, \mn@doi [A&A]
  {10.1051/0004-6361:200500127}, 437, L19–L22

\bibitem[\protect\citeauthoryear{Owers, Randall, Nulsen, Couch, David  \&
  Kempner}{Owers et~al.}{2011}]{Owers_2011}
Owers M.~S.,  Randall S.~W.,  Nulsen P. E.~J.,  Couch W.~J.,  David L.~P.,
  Kempner J.~C.,  2011, \mn@doi [ApJ] {10.1088/0004-637x/728/1/27}, 728, 27

\bibitem[\protect\citeauthoryear{Parkash, Brown, Jarrett  \& Bonne}{Parkash
  et~al.}{2018}]{Parkash_2018}
Parkash V.,  Brown M. J.~I.,  Jarrett T.~H.,   Bonne N.~J.,  2018, \mn@doi
  [ApJ] {10.3847/1538-4357/aad3b9}, 864, 40

\bibitem[\protect\citeauthoryear{{Planck Collaboration} et~al.,}{{Planck
  Collaboration} et~al.}{2015}]{Planckcollab_2016}
{Planck Collaboration} et~al., 2015, \mn@doi [A&A]
  {10.1051/0004-6361/201525830}, 594, A13

\bibitem[\protect\citeauthoryear{Poggianti, Bridges, Komiyama, Yagi, Carter,
  Mobasher, Okamura  \& Kashikawa}{Poggianti et~al.}{2004}]{Poggianti_2004}
Poggianti B.~M.,  Bridges T.~J.,  Komiyama Y.,  Yagi M.,  Carter D.,  Mobasher
  B.,  Okamura S.,   Kashikawa N.,  2004, \mn@doi [ApJ] {10.1086/380195}, 601,
  197–213

\bibitem[\protect\citeauthoryear{Péroux \& Howk}{Péroux \&
  Howk}{2020}]{P_roux_2020}
Péroux C.,  Howk J.~C.,  2020, \mn@doi [\araa]
  {10.1146/annurev-astro-021820-120014}, 58, 363–406

\bibitem[\protect\citeauthoryear{Ramatsoku et~al.,}{Ramatsoku
  et~al.}{2020}]{Ramatsoku_2020}
Ramatsoku M.,  et~al., 2020, \mn@doi [A\&A] {10.1051/0004-6361/202037759}, 640,
  A22

\bibitem[\protect\citeauthoryear{Rhee, Zwaan, Briggs, Chengalur, Lah, Oosterloo
   \& Hulst}{Rhee et~al.}{2013}]{Rhee_2013}
Rhee J.,  Zwaan M.~A.,  Briggs F.~H.,  Chengalur J.~N.,  Lah P.,  Oosterloo T.,
    Hulst T. v.~d.,  2013, \mn@doi [MNRAS] {10.1093/mnras/stt1481}, 435,
  2693–2706

\bibitem[\protect\citeauthoryear{Richard, Kneib, Limousin, Edge  \&
  Jullo}{Richard et~al.}{2009}]{Richard_2009}
Richard J.,  Kneib J.-P.,  Limousin M.,  Edge A.,   Jullo E.,  2009, \mn@doi
  [MNRAS: Letters] {10.1111/j.1745-3933.2009.00796.x}, 402, L44–L48

\bibitem[\protect\citeauthoryear{Roediger \& Brüggen}{Roediger \&
  Brüggen}{2007}]{Roediger_2007}
Roediger E.,  Brüggen M.,  2007, \mn@doi [MNRAS]
  {10.1111/j.1365-2966.2007.12241.x}, 380, 1399–1408

\bibitem[\protect\citeauthoryear{Salerno et~al.,}{Salerno
  et~al.}{2020}]{Salerno_2020}
Salerno J.~M.,  et~al., 2020, \mn@doi [MNRAS] {10.1093/mnras/staa545}, 493,
  4950–4959

\bibitem[\protect\citeauthoryear{Saponara, Koribalski, Benaglia  \&
  Fernández~López}{Saponara et~al.}{2017}]{Saponara_2017}
Saponara J.,  Koribalski B.~S.,  Benaglia P.,   Fernández~López M.,  2017,
  \mn@doi [MNRAS] {10.1093/mnras/stx2475}, 473, 3358–3366

\bibitem[\protect\citeauthoryear{Serra et~al.,}{Serra
  et~al.}{2012}]{Serra_2012b}
Serra P.,  et~al., 2012, \mn@doi [MNRAS] {10.1111/j.1365-2966.2012.20219.x},
  422, 1835–1862

\bibitem[\protect\citeauthoryear{Serra et~al.,}{Serra
  et~al.}{2015}]{Serra_2015}
Serra P.,  et~al., 2015, \mn@doi [MNRAS] {10.1093/mnras/stv079}, 448,
  1922–1929

\bibitem[\protect\citeauthoryear{Shipley et~al.,}{Shipley
  et~al.}{2018}]{Shipley_2018}
Shipley H.~V.,  et~al., 2018, \mn@doi [ApJ Supplement Series]
  {10.3847/1538-4365/aaacce}, 235, 14

\bibitem[\protect\citeauthoryear{Solanes, Manrique, García‐Gómez,
  González‐Casado, Giovanelli  \& Haynes}{Solanes
  et~al.}{2001}]{Solanes_2001}
Solanes J.,  Manrique A.,  García‐Gómez C.,  González‐Casado G.,
  Giovanelli R.,   Haynes M.,  2001, \mn@doi [ApJ] {10.1086/318672}, 548,
  97–113

\bibitem[\protect\citeauthoryear{{Sorgho}, {Hess}, {Carignan}  \&
  {Oosterloo}}{{Sorgho} et~al.}{2017}]{Sorgho_2017}
{Sorgho} A.,  {Hess} K.,  {Carignan} C.,   {Oosterloo} T.~A.,  2017, \mn@doi
  [\mnras] {10.1093/mnras/stw2341}, \href
  {https://ui.adsabs.harvard.edu/abs/2017MNRAS.464..530S} {464, 530}

\bibitem[\protect\citeauthoryear{Verheijen}{Verheijen}{2004}]{Verheijen_2004}
Verheijen M. A.~W.,  2004, \mn@doi [Proceedings of the International
  Astronomical Union] {10.1017/s1743921304000833}, 2004

\bibitem[\protect\citeauthoryear{Verheijen, van Gorkom, Szomoru, Dwarakanath,
  Poggianti  \& Schiminovich}{Verheijen et~al.}{2007}]{Verheijen_2007}
Verheijen M.,  van Gorkom J.~H.,  Szomoru A.,  Dwarakanath K.~S.,  Poggianti
  B.~M.,   Schiminovich D.,  2007, \mn@doi [ApJ] {10.1086/522621}, 668,
  L9–L13

\bibitem[\protect\citeauthoryear{Vollmer, Soida, Beck, Chung, Urbanik, Chyży,
  Otmianowska-Mazur  \& Kenney}{Vollmer et~al.}{2013}]{Vollmer_2013}
Vollmer B.,  Soida M.,  Beck R.,  Chung A.,  Urbanik M.,  Chyży K.~T.,
  Otmianowska-Mazur K.,   Kenney J. D.~P.,  2013, \mn@doi [A&A]
  {10.1051/0004-6361/201321163}, 553, A116

\bibitem[\protect\citeauthoryear{Wang, Koribalski, Serra, van~der Hulst,
  Roychowdhury, Kamphuis  \& N.~Chengalur}{Wang et~al.}{2016}]{Wang_2016}
Wang J.,  Koribalski B.~S.,  Serra P.,  van~der Hulst T.,  Roychowdhury S.,
  Kamphuis P.,   N.~Chengalur J.,  2016, \mn@doi [MNRAS]
  {10.1093/mnras/stw1099}, 460, 2143–2151

\bibitem[\protect\citeauthoryear{Wetzel, Tinker, Conroy  \& van~den
  Bosch}{Wetzel et~al.}{2013}]{Wetzel_2013}
Wetzel A.~R.,  Tinker J.~L.,  Conroy C.,   van~den Bosch F.~C.,  2013, \mn@doi
  [MNRAS] {10.1093/mnras/stt469}, 432, 336–358

\bibitem[\protect\citeauthoryear{Williamson et~al.,}{Williamson
  et~al.}{2011}]{Williamson_2011}
Williamson R.,  et~al., 2011, \mn@doi [ApJ] {10.1088/0004-637x/738/2/139}, 738,
  139

\bibitem[\protect\citeauthoryear{van~der Wel, Bell, Holden, Skibba  \&
  Rix}{van~der Wel et~al.}{2010}]{van_der_Wel_2010}
van~der Wel A.,  Bell E.~F.,  Holden B.~P.,  Skibba R.~A.,   Rix H.-W.,  2010,
  \mn@doi [ApJ] {10.1088/0004-637x/714/2/1779}, 714, 1779–1788

\makeatother
\end{thebibliography}
\bsp	
\label{lastpage}
\end{document}